\newif\ifExtended\Extendedtrue

\documentclass[letterpaper,twocolumn,10pt]{article}
\usepackage{usenix2019_v3}

\usepackage[square,comma,numbers,sort&compress]{natbib}
\usepackage{listings}
\usepackage{xcolor}
\usepackage{tabularx}
\usepackage[frozencache,cachedir=.]{minted}
\usepackage{inconsolata}
\usepackage{algorithm}
\usepackage{nameref}
\usepackage{graphicx}
\usepackage{times}
\usepackage{xspace}
\usepackage{microtype}
\usepackage{url}
\usepackage{booktabs}
\usepackage{amssymb}
\usepackage{enumitem}
\usepackage[export]{adjustbox}
\usepackage{pbox}
\usepackage{multirow}
\usepackage{titling}
\usepackage[all=normal,paragraphs=tight,floats=tight]{savetrees}
\usepackage[compact,nonindentfirst]{titlesec}

\newcommand{\ignore}[1]{}

\newcommand{\para}[1]{\smallskip\noindent\textbf{{#1.}\xspace}}

\newcommand{\vtainted}{\code{tainted_volatile}\xspace}

\newcommand{\tainted}{\code{tainted}\xspace}
\newcommand{\taintedW}[1]{\code{tainted<#1>}\xspace}

\newcommand{\unsafe}{\code{unsafeUnverified()}\xspace}

\newcommand\copyAndVerify{\code{copyAndVerify()}\xspace}
\newcommand\verify{\code{verify()}\xspace}
\newcommand\sys{RLBox\xspace}

\newcommand\libjpeg{\textsf{libjpeg}\xspace}
\newcommand\jpeg{\libjpeg}
\newcommand\turbo{\textsf{libjpeg-turbo}\xspace}
\newcommand\libpng{\textsf{libpng}\xspace}
\newcommand\png{\libpng}
\newcommand\zlib{\textsf{zlib}\xspace}
\newcommand\graphite{\textsf{libGraphite}\xspace}
\newcommand\ff{Firefox\xspace}
\newcommand\libvorbis{\textsf{libvorbis}\xspace}
\newcommand\vorbis{\libvorbis}
\newcommand\libtheora{\textsf{libtheora}\xspace}
\newcommand\theora{\libtheora}
\newcommand\libvpx{\textsf{libvpx}\xspace}
\newcommand\vpx{\libvpx}

\newcommand\libmarkdown{\textsf{libmarkdown}\xspace}
\newcommand\modmarkdown{\textsf{mod\_markdown}\xspace}

\newcommand\unmodffpre{stock\xspace}

\newcommand\unmodff{\unmodffpre Firefox\xspace}

\def\dash---{\kern.16667em---\penalty\exhyphenpenalty\hskip.16667em\relax}

\fvset{fontsize=\footnotesize, xleftmargin=5pt, numbersep=5pt}

\usepackage{subcaption}
\usepackage[font=small,labelfont=bf]{caption}
\def\C++{
C\kern-.1667em\raise.30ex\hbox{\smaller{++}}%
\spacefactor1000
}

\definecolor{xxxcolor}{rgb}{0.8,0,0}

\newcommand{\code}[1]{\mintinline[breaklines, breakafter= , 
fontsize=\small]{cpp}{#1}}

\newcommand{\footnotecode}[1]{\mintinline[fontsize=\footnotesize]{cpp}{#1}}

\newenvironment{CompactItemize}%
  {\begin{list}{$\blacktriangleright$}%
    {\leftmargin=\parindent \itemsep=2pt \topsep=2pt
     \parsep=0pt \partopsep=0pt}}%
  {\end{list}}

\newenvironment{CompactEnumerate}{
\begin{enumerate}[leftmargin=*]
}{\end{enumerate}}

\date{\vspace{-1em}}
\hypersetup{breaklinks=true}

\usepackage[firstpage]{draftwatermark}
\SetWatermarkText{\hspace*{6in}\raisebox{9.5in}{\includegraphics{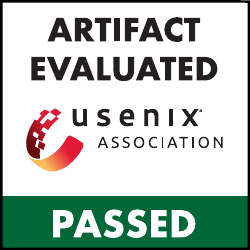}}}
\SetWatermarkAngle{0}

\ifExtended
\newcommand{\appref}[2]{Appendix~\ref{#2}\xspace}
\else
\newcommand{\appref}[2]{Appendix~{#1} of~\cite{extended}\xspace}
\fi

\ifExtended
\else
\fi
\begin{document}

\setlength{\droptitle}{-2em}

\ifExtended
\title{Retrofitting Fine Grain Isolation in the Firefox Renderer\\\large{Extended Version} \vspace{-1em}}
\else
\title{Retrofitting Fine Grain Isolation in the Firefox Renderer\vspace{-1.2em}}
\fi

\newcommand{\ucsdmark}{{$^{\dagger}$}\xspace}
\newcommand{\stanmark}{{$^{\ast}$}}
\newcommand{\mozmark}{{$^{\diamond}$}\xspace}
\newcommand{\utmark}{{$^{\star}$}\xspace}

\author{
    Shravan Narayan\ucsdmark
    \quad
    Craig Disselkoen\ucsdmark
    \quad
    Tal Garfinkel\stanmark
    \quad
    Nathan Froyd\mozmark
    \quad
    Eric Rahm\mozmark
    \\
    Sorin Lerner\ucsdmark
    \quad
    Hovav Shacham$^{\star,\dagger}$
    \quad
    Deian Stefan\ucsdmark
    \\
    \small{
      \ucsdmark{UC San Diego}\qquad
      \stanmark{Stanford}\qquad
      \mozmark{Mozilla}\qquad
      \utmark{UT Austin}\qquad
    }
} 

\maketitle

\begin{abstract}

\ff and other major browsers rely on dozens of third-party libraries to
render audio, video, images, and other content. These libraries are a
frequent source of vulnerabilities. To mitigate this threat, we 
are migrating \ff to an architecture that isolates these libraries in
lightweight sandboxes, dramatically reducing the impact of a compromise.

Retrofitting isolation can be labor-intensive, very prone to security bugs, and
requires critical attention to performance.  To help, we developed~\sys, a
framework that minimizes the burden of converting \ff to securely and
efficiently use untrusted code. To enable this, \sys employs static information
flow enforcement, and lightweight dynamic checks, expressed directly in the
C++ type system.

\sys supports efficient sandboxing through either software-based-fault
isolation or multi-core process isolation. Performance overheads are modest
and transient, and have only minor impact on page latency.  We demonstrate
this by sandboxing performance-sensitive image decoding libraries (\libjpeg
and \libpng), video decoding libraries (\libtheora and \libvpx), the
\libvorbis audio decoding library, and the \zlib decompression library.

\sys, using a WebAssembly sandbox, has been integrated into production \ff to
sandbox the \graphite font shaping library.
\end{abstract}

\section{Introduction}
\label{sec:intro}

All major browsers today employ coarse grain {\it privilege separation} to
limit the impact of vulnerabilities.
To wit, they run \emph{renderers}---the portion of the browser that handles
untrusted user content from HTML parsing, to JavaScript execution, to image
decoding and rendering---in separate sandboxed
processes~\cite{barth-et-al:chromium:08, mozilla-sandbox, reis-isolating}.
This stops web attackers that manage to compromise the renderer from abusing
local OS resources to, say, install malware.

Unfortunately, this is no longer enough: nearly everything we care about today
is done through a website.
By compromising the renderer, an attacker gets total control of the current
site and, often, any other sites the browser has credentials for~\cite{UXSS}.
With services like Dropbox and Google Drive, privilege separation is
insufficient even to protect local files that sync with the
cloud~\cite{jia-et-al:weblocal:ccs16}.

Browser vendors spend a huge amount of engineering effort trying to find
renderer vulnerabilities in their own code~\cite{ffjsfuzz}.  Unfortunately, many
remain---frequently in the dozens of third-party libraries used by the renderer
to decode audio, images, fonts, and other content.
For example, an out-of-bounds write in \libvorbis was used to exploit Firefox
at Pwn2Own 2018~\cite{pwn2own2018}.
Both Chrome and Firefox were vulnerable to an integer-overflow bug in the \vpx
video decoding library~\cite{vpxbug}.
Both also rely on the Skia graphics library, which had four remote code
execution bugs until recently~\cite{skia-rce-android,
skia-rce-1}.

To appreciate the impact of these vulnerabilities and the difficulty of
mitigating them, consider a typical web user, Alice, that uses Gmail to read
email in her browser.
Suppose an intruder, Trudy, sends Alice an email that contains a link to her
malicious site, hosted on \texttt{sites.google.com}.
If Alice clicks on the link, her browser will navigate her to Trudy's site, which
can embed an \texttt{.ogg} audio track or \texttt{.webm} video to 
exploit vulnerabilities in \vorbis and \vpx and compromise
the renderer of Alice's browser.  Trudy now has total control of Alice's Gmail
account. Trudy can read and send emails as Alice, for example, to respond to
password reset requests from other sites Alice belongs to.  In most cases,
Trudy can also attack cross site~\cite{UXSS}, i.e., she can access any other
site that Alice is logged into~(e.g., Alice's {\tt amazon.com} account).

Recent version of Chrome (and upcoming versions of Firefox) support {\it
Site Isolation}~\cite{site-isolation-usenix}, which
isolates different
\emph{sites} from each other~(e.g., \texttt{*.google.com} from
\texttt{*.amazon.com}) to prevent such cross-site attacks.
Unfortunately, Trudy might still
be able to access \texttt{\{drive,pay,cloud\}.google.com}, which manage Alice's
files, online payments, and cloud infrastructure---since the renderer that loads
the malicious \texttt{.ogg} and \texttt{.webm} content might still be
running in the same process as those origins.

For many sites, Trudy might not even need to upload malicious content to the
(trusted) victim origin (\texttt{sites.google.com} in our example).
Most web applications load content, including images, fonts, and video, from
different origins.
Of the Alexa top 500 websites, for example, over 93\% of the sites load at least
one such cross-origin resource~(\S\ref{subsec:cross_origin}).
And the libraries handling such content are not isolated from the embedding
origin, even with Site Isolation~\cite{site-isolation-usenix}.

To mitigate these vulnerabilities, we need to harden the renderer itself.  To this end, we extend the \ff renderer to
isolate
third party libraries in fine grain sandboxes.  Using this, we can prevent a
compromised library from gaining control of the current origin or any other
origin in the browser.

Making this practical poses three significant challenges across three
dimensions.
First, {\it engineering effort}---we need to minimize the upfront work required
to change the renderer to use sandboxing, especially as this is multiplied
across dozens of libraries;
minimizing changes to libraries is also important as this can
significantly increase the burden of tracking upstream changes.
Second, {\it security}---the renderer was not built
to protect itself from libraries; thus, we have to sanitize all data and regulate control
flow between the library and renderer to prevent libraries from breaking out of
the sandbox.
In our experience, bugs at the library-renderer boundary are not only easy to
overlook, but can nullify any sandboxing effort---and other developers,
not just us, must be able to securely sandbox new libraries.
Finally, {\it efficiency}---the renderer is performance critical, so adding
user-visible latency is not acceptable.

To help us address these challenges, we develop a framework called \sys that
makes data- and control-flow at the library-renderer interface explicit, using
types. Unlike prior approaches to sandbox automation that rely on extensive
custom analysis frameworks~(\S\ref{sec:related}), \sys is simply a
library\footnote{
Our only external tooling is a \textasciitilde{}100LOC Clang
plugin, described in Section~\ref{subsec:impl_taint}, that makes up for C++'s
currently limited  support for reflection on structs.
}
that leverages the C++ type system and is easy to incorporate into \ff's
predominantly C++ codebase.

Using type information, \sys can identify where security checks are needed,
automatically insert dynamic checks when possible, and force compiler errors
for any security checks that require additional user intervention.  Our
type-driven approach enables a systematic way to migrate \ff's renderer to use
sandboxed libraries and allows \sys to support secure and
efficient sharing of data structures between the renderer and library (e.g., by
making shared memory operations safe and by lazily copying data out of the sandbox).

To enable efficient sandboxing, we adapt and evaluate two isolation
mechanisms for library sandboxing: software-based fault isolation (SFI)
leveraging Google's Native Client (NaCl)~\cite{nacl,nacl-amd64} and a
multi-core process-based approach.
We also explore applying sandboxing at different granularities (e.g.,
per-origin and per-library sandboxing) to find the appropriate balance between
security and sandboxing overhead.

To evaluate \sys, we sandbox several libraries in Firefox: the \jpeg and \png
image decoding libraries, the \vpx and \theora video decoding libraries, the
\vorbis audio decoding library, and the \zlib decompression library.
Browsing a representative sample of both popular and unpopular websites
(\S\ref{sec:eval}), we find the end-to-end memory overhead of \sys to be
modest---25\% with SFI, 18\% with process isolation---and transient,
appearing only at content load time.
The impact on page latency is small: 3\% and 13\% with SFI
and process isolation, respectively.
Our sandboxing does not noticeably impact the video frame rates nor audio
decoding bitrate.

Our evaluation shows that retrofitting fine grain isolation, especially
using SFI, is practical---and we've been integrating \sys into production
Firefox~\cite{ff-integration}.\footnote{The Tor team is integrating our patches
into the Tor Browser~\cite{tor-integration}.}
Since NaCl has been deprecated~\cite{nacl-deprecate} in favor of WebAssembly
(Wasm)~\cite{wasm}, our production sandbox also uses Wasm.
We used \sys with this Wasm-based sandbox to isolate the \graphite font shaping
library and are in the process of migrating several others~\cite{moz-blog,
ff-integration}.
We describe this effort in Section~\ref{sec:upstream}.

Though we developed \sys to sandbox libraries in Firefox, \sys is a general
library-sandboxing framework that can be used outside Firefox.
To demonstrate this, we use \sys to sandbox libraries in two different
contexts: the Apache web server and Node.js runtime.
For Apache, we sandbox the \libmarkdown library that is
used in the \modmarkdown module~\cite{mod_markdown};
we find that \sys with the SFI sandbox increases the tail latency of the
Apache server by 10\% (4ms) and decreases the throughput by 27\% (256
requests/second).
For Node.js, we sandbox the C \textsf{bcrypt} library that is used by the
JavaScript \textsf{bcrypt} module~\cite{node_bcrypt}; we measure \sys with SFI
to impose an overhead of 27\% on hashing throughput.

\para{Contributions}
We present the case for sandboxing third party libraries in the browser
renderer, and potential architectural trade-offs, including our
approach~(\S\ref{sec:fine_grain}).
We offer a taxonomy of security pitfalls encountered while migrating the
\ff code base to this architecture that were largely overlooked by previous
work~(\S\ref{sec:problems}), and \sys, a
framework we developed to prevent these pitfalls that leverages the
C++ type system to enforce safe data and control flow~(\S\ref{sec:design}), and
enables an incremental compiler-driven approach to migrating code to a
sandboxed architecture~(\S\ref{sec:migration}).
We describe our implementation, including our software fault
isolation and multi-core process-based isolation mechanisms~(\S\ref{sec:impl}),
and evaluate the performance of \sys~(\S\ref{sec:eval}).
We close with a discussion of related work~(\S\ref{sec:related}) and our
effort upstreaming \sys into production \ff~(\S\ref{sec:upstream}).

\para{Availability}
All work presented in this paper, including our modified \ff builds, the \sys 
library, and benchmarks are available and open source.\footnote{
Available at: \url{https://usenix2020-aec.rlbox.dev}.
}

\section{Fine grain sandboxing: how and why}
\label{sec:fine_grain}

Renderers rely on dozens of third-party libraries to support media decoding 
and other tasks (e.g., decompression, which sites use to optimize
page load times and bandwidth consumption).
These are written almost exclusively in C and tasked with parsing a wide
range of complex inputs. Unsurprisingly, exploitable vulnerabilities in this
code are relatively frequent, even after years of scrutiny.

These libraries are a compelling place to employ sandboxing inside the renderer
for several reasons.  First, media content such as images and video are rich
attack vectors, as web applications allow them to be shared pervasively.
Over 93\% of the Alexa Top 500 websites load such content cross-origin
(\S\ref{subsec:cross_origin}).
And nearly all forms of social media and peer-to-peer messaging platforms
enable the sharing of images and video.

Next, content libraries can be effectively sandboxed, as they require little
privilege to operate, i.e., once these libraries are memory isolated, the harm
they can inflict is minimal.  For example, an attacker that compromises an
image decoding library could at worst change how images display.
In contrast, sandboxing a highly privileged component like the JavaScript
engine is largely ineffectual. An attacker with control over the JavaScript
engine can run arbitrary JavaScript code and thus already has complete
control of the web application.

Finally, the existing library-renderer interface provides a natural place to
partition code.  Compared to coarse grain techniques like privilege
separation or Site Isolation, which spin up entire new renderer processes,
spinning up a sandbox for a library is very cheap~(\S\ref{sec:eval}).
Moreover, because library sandboxes are only needed during content decoding,
their memory overhead is transient.

\para{Isolation strategies}
A key question remains: \emph{what grain of isolation should be employed?} In 
particular, different architectures have different implications for performance
and security.
Prior to \sys, \ff was largely exploring a coarse grain approach to library
sandboxing, placing certain media libraries into a single
sandboxed media process~\cite{mozilla-sandbox}.
This approach has some benefits for performance as there is only one sandbox,
but trades off security.

First, the assurance of the sandbox is reduced to that of the weakest library.
This is less than ideal, especially when we consider the long tail of
infrequently used media libraries required to preserve web compatibility (e.g.,
Theora) which often contain bugs.
Next, the attacker gains the power of the most capable library.
Some libraries handle {\it active content}---\zlib, for example,
is used to decompress HTTP requests that could contain HTML or JavaScript---as
opposed to {\it passive content} such as images or fonts.
Thus compromising a passive library like \vorbis, could still enable
powerful attacks---e.g., modify the JavaScript decompressed by \zlib. 
When multiple renderers share a common library sandbox, an intruder can
attack across tabs, browsing profiles, or sites.
Finally, coarse grain sandboxing does not scale to highly performance-critical
libraries, such as \jpeg and \png~(\S\ref{subsubsec:images}).

\sys lets us employ more granular sandboxing policies that can address these
shortcomings.
Its flexibility lets us explore the performance implications
of various sandboxing architectures with different isolation
mechanisms~(\S\ref{subsec:end2end}).
 
In this paper, we largely employ a unique sandbox per \code{<renderer, library,
content-origin, content-type>}. This mitigates many of the problems noted
above, while still offering modest memory overheads.
Per-renderer sandboxing prevents attacks across tabs and browsing profiles.
Per-library ensures that a weakness in one library does not impact any other
library.
Per-content-origin sandboxing prevents cross origin (and thus cross
site) attacks on content.
For example, a compromise on \texttt{sites.google.com} as discussed
in our example in Section~\ref{sec:intro}, should not impact content from
\texttt{pay.google.com}. Per-content-type sandboxing addresses the problem of
passive content influencing active content.

Both finer and coarser grain policies are practically useful, though.
In production \ff, for example, we create a fresh sandbox for each Graphite
font instance~(\S\ref{sec:upstream}).
But, we also foresee libraries where, say, same-origin is sufficient.

\para{Attacker model}
We assume a web attacker that serves malicious (but passive) content---from an
origin they control or by uploading the content to a trusted origin---which
leads to code execution (e.g., via a memory safety vulnerability) in a \sys
sandbox.
\sys ensures that such an attacker can only affect (corrupt the rendering of
and potentially leak) content of the same type, from the same origin.
Per-object (or per-instance) sandboxing can further reduce the damage of such
attacks. We, however, only use this policy when sandboxing audio, videos, and
font shaping---we found the overheads of doing this for images to be
prohibitive.

We consider side channels out-of-scope, orthogonal challenges.
With side channels, an attacker doesn't need to exploit renderer 
vulnerabilities to learn cross-origin information, as browsers like \ff largely
do not prevent cross-origin leaks via side channels.
FuzzyFox~\cite{fuzzyfox} and cross-origin read
blocking~\cite{site-isolation-usenix} are promising ways to tackle these
channels.

For the same reason, we consider transient execution attacks (e.g.,
Spectre~\cite{spectre}) out of scope.
We believe that our SFI and our process-based isolation mechanisms make many
of these attacks harder to carry out---e.g., by limiting transient reads and
control flow to the sandbox memory and code, respectively---much like Site
Isolation~\cite{site-isolation-usenix}.
But, in general, this is not enough: an attacker could potentially exploit code
in the renderer to transiently leak sensitive data.
We leave the design of a Spectre-robust sandbox to future work.
 
The protections offered by \sys are only valid if the \ff code that interfaces
with the sandboxed library code is retrofitted to account for untrusted code
running in the library sandbox.
As we discuss next, this is usually notoriously difficult to get right.
\sys precisely reduces this burden to writing a series of validation functions.
In our attacker model, we assume these functions to be correct.

\section{Pitfalls of retrofitting protection}
\label{sec:problems}

The \ff renderer was written assuming libraries are trusted. To benefit from
sandboxing requires changing our threat model to assume libraries are
untrusted, and modify the renderer-library interface accordingly (e.g, to
sanitize untrusted inputs).

While migrating to this model we made numerous mistakes---overlooking attack
vectors and discovering many bugs only after building \sys to help detect them.
We present a brief taxonomy of these mistakes, with examples drawn from the
code snippet illustrating the interface between the renderer's JPEG decoder and
\jpeg\footnote{We use \jpeg interchangeably with \turbo, the faster fork of the
original \jpeg library which is used in \ff.}  shown in
Figure~\ref{fig:jpeg_decoder}. We discuss how \sys helps prevent these attacks
in Section~\ref{sec:design}.

For the rest of this section, we assume that \jpeg is fully sandboxed or memory 
isolated, i.e., \jpeg code is restricted from accessing arbitrary memory 
locations in the renderer process, and may only access memory explicitly 
dedicated to the sandbox---the \emph{sandbox memory}. The renderer itself can 
access any memory location including sandbox memory.

\begin{figure}
\begin{minted}[linenos=true, breaklines=true, mathescape=true,
escapeinside=||]{cpp}
// InitInternal() registers callbacks for libjpeg to call while decoding an image
nsresult nsJPEGDecoder::InitInternal() {
  ...
  mInfo.client_data = (void*)this;|\label{line:cd_app_ptr_write}|
  ...
  //Callbacks invoked by libjpeg
  mErr.pub.error_exit = my_error_exit;
  mSourceMgr.fill_input_buffer = fill_input_buffer;|\label{line:cd_register_cb}|
  mSourceMgr.skip_input_data = skip_input_data;
  ...
}

// Adjust output buffers for decoded pixels
void nsJPEGDecoder::OutputScanlines(...) {
  ...
  while (mInfo.output_scanline < mInfo.output_height) {|\label{line:cd_check}|
    ...
    imageRow = ... +  (mInfo.output_scanline * mInfo.output_width);|\label{line:cd_use}|
    ...
  }
  ...
}

// Invoked if some input bytes are not needed
void skip_input_data (..., long num_bytes) {|\label{line:cd_num_bytes}|
  ...
  if (num_bytes > (long)src->bytes_in_buffer) { |\label{line:cd_branch}|
    ...
  } else {
    src->next_input_byte += num_bytes;
  }
}

// Invoked repeatedly to get input as it arrives
void fill_input_buffer (j_decompress_ptr jd) {|\label{line:cd_jd}|
  struct jpeg_source_mgr* src = jd->src;|\label{line:cd_jdsrc}|
  nsJPEGDecoder* decoder = jd->client_data;|\label{line:cd_client_data_decoder}|
  ...
  src->next_input_byte = new_buffer;|\label{line:cd_next_input_byte_write}|
  ...
  if (/* buffer is too small */) {
    JOCTET* buf = (JOCTET*) realloc(...);|\label{line:cd_realloc}|
    if (!buf) {
      decoder->mInfo.err->msg_code = JERR_OUT_OF_MEMORY;|\label{line:cd_get_unsandboxed}|
      ...
    }
    ...
  }
  ...
  memmove(decoder->mBackBuffer + decoder->mBackBufferLen, src->next_input_byte, src->bytes_in_buffer);|\label{line:cd_usesrc}|
  ...
}

// Invoked on a decoding error
void my_error_exit (j_common_ptr cinfo) {|\label{line:cd_my_error_exit}|
  decoder_error_mgr* err = cinfo->err;
  ...
  longjmp(err->setjmp_buffer, error_code);|\label{line:cd_longjmp}|
}
\end{minted}
\caption{
 {\it The renderer-library interface:} this code snippet illustrates the
 renderer's interface to the JPEG decoder and is used as a running example.
 The decoder uses \jpeg's streaming interface to decode images one pixel-row at
 a time, as they are received.  Receiving and decoding concurrently is
 critical for responsiveness.
}
\label{fig:jpeg_decoder}
\end{figure}

\subsection{Insecure data flow}

\para{Failing to sanitize data} Failing to sanitize data received from \jpeg
including function return values, callback parameters, and data read from
sandbox shared memory can leave the renderer vulnerable to attack.  For
example, if the renderer uses the \code{num_bytes} parameter to the
\code{skip_input_data()} callback on line~\ref{line:cd_num_bytes} of
Figure~\ref{fig:jpeg_decoder} without bounds checking it, an
attacker-controlled \jpeg could could force it to overflow or underflow the
\code{src->next_input_byte} buffer.

Pointer data is particularly prone to attack, either when pointers are used
directly (with C++'s \code{*} and \code{->} operators) or indirectly (via memory
functions such as \code{memcpy()} and \code{memmove()}, array indexing
operations, etc.).
For example, if the parameter \code{jd} of \code{fill_input_buffer()} is not
sanitized (line~\ref{line:cd_jd}), the read of \code{jd->src} on
line~\ref{line:cd_jdsrc} becomes an arbitrary-read gadget.
In the same callback, if both \code{jd} and \code{src} are unsanitized, the 
write to  \code{src->next_input_byte} on 
line~\ref{line:cd_next_input_byte_write} becomes an arbitrary-write gadget.
Similar attacks using the \code{memmove()} on line~\ref{line:cd_usesrc} are
possible.

\para{Missing pointer swizzles}
Some sandboxing mechanisms---e.g., NaCl~(\S\ref{subsec:isol_mechanisms}) and
Wasm~(\S\ref{sec:upstream})---use alternate pointer representations.
Some sandboxing tools e.g., NaCl (\S\ref{subsec:isol_mechanisms}) and Wasm 
(\S\ref{sec:upstream}) use alternate pointer representations for efficiency.

Though this is often done for performance, in Wasm's case this is more
fundamental: Wasm pointers are 32-bit whereas Firefox pointers are 64-bit.
We must translate or \emph{swizzle} pointers to and from these alternate 
representations when data is transferred between the renderer and the sandboxed 
\jpeg.
 
We found that doing this manually is both tedious and extremely error prone.
This is largely because pointers can be buried many levels deep in nested data
structures.
Overlooking a swizzle either breaks things outright, or worse, silently 
introduces vulnerabilities.
For instance, failing to swizzle the nested pointer \code{mInfo.err} on
line~\ref{line:cd_get_unsandboxed} prior to dereferencing, can result in 
a write gadget (whose write-range depends on the precise pointer
representation).

\para{Leaking pointers}
Leaking pointers from the \ff renderer to the sandboxed \jpeg can allow an 
attacker to derandomize ASLR~\cite{shacham-aslr} or otherwise learn locations 
of code pointers (e.g., C++ virtual tables).  Together with an arbitrary-write 
gadget, this can allow an attacker-controlled \jpeg to execute arbitrary code 
in the renderer.

In our example, the renderer saves pointers to \code{nsJPEGDecoder} objects in
\jpeg \code{struct}s (line~\ref{line:cd_app_ptr_write}), which alone allows
an attacker to locate code pointers---the \code{nsJPEGDecoder} class is derived
from the \code{Decoder} class, which defines virtual methods and thus has a
virtual table pointer as the first field. 
Even initializing callbacks (line \ref{line:cd_register_cb}) 
could leak pointers to functions and reveal the location 
of \ff's code segment\footnote{Whether callback locations are leaked depends on
the underlying sandboxing mechanism. While both our process isolation and NaCl
use jump tables and thus do not leak, other sandbox implementations could
leak such information.}.

\para{Double fetch bugs}
\sys uses shared memory~(\S\ref{sec:design}) to efficiently marshal objects
between the renderer and the sandboxed libraries.
This, unfortunately, introduces the possibility of double fetch
bugs~\cite{double-fetch-static-analysis, double-fetch-hardware, 
xu2018precise}.
 
Consider the \code{mInfo} object used in Figure~\ref{fig:jpeg_decoder}.
Since this object is used by both \jpeg and the renderer,
\sys stores it in shared memory. Now consider the bounds check of 
\code{mInfo.output_scanline} on line~\ref{line:cd_check} prior to the 
assignment of output buffer \code{imageRow}.
In a concurrent \jpeg sandbox thread, an attacker can modify 
\code{mInfo.output_scanline} after the check (line~\ref{line:cd_check}), and
before the value is fetched (again) and used on line~\ref{line:cd_use}.
This would bypasses the bounds check, leading to an arbitrary-write gadget.
While this example is obvious, double-fetch bugs often span function boundaries
and are much harder to spot.

\subsection{Insecure control flow}

Isolation prevents arbitrary control transfers from the sandbox into the
renderer.
Thus, out of necessity, callbacks for \jpeg must be explicitly exposed.
But this alone is not sufficient to prevent attacks.

\para{Corrupted callback state}
Callbacks may save state in the sandboxed library.
An attacker-controlled \jpeg can abuse the control flow of callbacks by 
corrupting this state. 
For example, on line~\ref{line:cd_app_ptr_write} of
Figure~\ref{fig:jpeg_decoder}, the renderer stores a pointer to the
\code{nsJPEGDecoder} object into the \code{client_data} field of
\code{mInfo}. Inside \code{fill_input_buffer()}  this pointer is used to access
the \code{nsJPEGDecoder} object (line~\ref{line:cd_client_data_decoder}).
Failing to sanitize \code{client_data} before using it allows an attacker to
set the pointer to a maliciously crafted object and hijack control flow when 
\ff invokes a virtual method on this object.

\para{Unexpected callback invocation}
Security bugs can also occur if an attacker controlled \jpeg invokes a 
permitted callback at unexpected times.
Consider the \code{my_error_exit()} callback function, which uses \code{longjmp()} to implement
error handling. On line~\ref{line:cd_longjmp}, \code{longjmp()} changes the
instruction pointer of the renderer based on information stored in
\code{setjmp_buffer}. If an attacker invokes \code{my_error_exit()}
before \code{setjmp_buffer} is initialized, they can (again)
hijack the renderer control flow.

\para{Callback state exchange attacks}
Threading introduces another vector to attack callback state. 
When \ff decodes two images in parallel, two decoder threads make calls to 
\jpeg.  \ff expects \jpeg to invoke the \code{fill_input_buffer()} callback on 
each thread with the corresponding \code{nsJPEGDecoder} object.  But, an 
attacker could supply the same \code{nsJPEGDecoder} object to both threads when 
calling \code{fill_input_buffer()}.  If the first thread reallocates the source 
buffer (line~\ref{line:cd_realloc}), while the second thread is using it to get 
input bytes, this can induce a data race and use-after-free vulnerability in 
turn.

\section{\sys: automating secure sandboxing}
\label{sec:design}

Modifying the \ff JPEG decoder to guard against all the attacks
discussed in Section~\ref{sec:problems} requires substantial code additions and
modifications.
Doing this manually is extremely error prone.
Moreover, it also makes the code exceedingly fragile:
anyone making subsequent changes to the decoder must now have an intimate
knowledge of all the necessary checks to have any hope of getting it right.
Multiply this by the number of libraries \ff supports and number of
developers working on the renderer, and the problem becomes intractable.

We built the \sys framework to tackle these challenges.
\sys helps developers migrate and maintain code in the \ff renderer to safely
use sandboxed libraries.
We designed \sys with the following goals in mind:

\begin{CompactEnumerate}

\item {\it Automate security checks: } Adding security checks un-assisted
is labor intensive and error prone, as discussed~(\S\ref{sec:problems}).
However, most of the sandboxing pitfalls can be detected and prevented through
static checks, i.e., through compile-time errors indicating where code
needs to be manually changed for security, or eliminated with dynamic checks
and sanitizations (e.g., pointer swizzling and bounds checking).

\item {\it No library changes:} We want to avoid making changes to libraries.
When libraries come from third parties we do not necessarily understand
their internals, nor do we want to. Additionally, any changes we make to libraries
increases the effort required to track upstream changes.

\item {\it Simplify migration:} \ff uses dozens of libraries, with occasional
additions or replacements. Consequently, we want to minimize the per-library
effort of using \sys, and minimize changes to the \ff renderer source.
\end{CompactEnumerate}

\noindent In the rest of the section we give an overview of the \sys framework
and describe how \sys addresses the pitfalls of Section~\ref{sec:problems} while
preserving these goals.

\subsection{\sys overview}
\sys makes data and control flow at the renderer-sandbox interface explicit 
through its type system and APIs in order to mediate these flows and enforce 
security checks across the trust boundary.

\sys mediates \emph{data flow} with \tainted types that
impose a simple static information flow control (IFC) 
discipline~\cite{sabelfeld2003ifc}.
This ensures that sandbox data is validated before any potentially unsafe use.
It also prevents pointer leaks into the sandbox that could break ASLR.  

\sys mediates \emph{control flow} through a combination of \tainted types and 
API design.
Tainting, for example, allows \sys to prevent branching on \tainted values that
have not been validated.
API design, on the other hand, is used to restrict control transfers between
the renderer and sandbox.
For instance, the renderer must 
use \code{sandbox_invoke()} to invoke functions in the sandbox; any callback
into the renderer by the sandbox must first be registered by the renderer using
the \code{sandbox_callback(callback_fn)} API.

Mediating control and data flow allows \sys to:

\begin{CompactItemize}
\item
{\bf Automate security checks:} Swizzling operations, performing checks that ensure
sandbox-supplied pointers point to sandbox memory, and identifying
locations \emph{where} tainted data must be validated is done 
automatically.

\item {\bf Minimize renderer changes:} \tainted data validation is enforced
only when necessary. Thus, benign operations such as adding or assigning \tainted
values, or writing to tainted pointers (which are checked to ensure they point
into the sandbox at creation time) are allowed by \sys's type system.  This
eliminates needless changes to the renderer, while still ensuring safety.

\item {\bf Efficiently share data structures:} Static checks ensure
that shared data is allocated in sandbox memory and accessed via \tainted
types.
Data structures received by the renderer sandbox are marshaled lazily;
this allows code to access a single field in a shared \code{struct} without
serializing a big object graph.
Finally, \sys provides helper APIs to mitigate double 
fetches~\cite{double-fetch-static-analysis, double-fetch-hardware} when 
accessing this data.

\item {\bf Assist with code migration:} Compile-time type and interface errors~(\S\ref{sec:migration}) guide the developer through the process of migrating a
library into a sandbox. Each compile error points to the next required code 
change---e.g., data that needs to be validated before use, or control transfer 
code that needs to be changed to use \sys APIs.

\item {\bf Bridge machine models:} Sandboxing mechanisms can have a
different machine model from the application (e.g., both Native Client and
WebAssembly use 32-bit pointers and 32-bit \code{long}s regardless of the
platform); by intercepting all data and control flow we can also automatically
translate between the application and sandbox machine models---and for Wasm we
do this~(\S\ref{sec:upstream}).

\end{CompactItemize}
\noindent
In the rest of this section, we discuss how tainting is used to mediate data
flow~(\S\ref{subsec:dataflow}) and control flow~(\S\ref{subsec:controlflow}).
We then describe how the renderer can validate and untaint
data~(\S\ref{subsec:dataflow_validation}). We detail our
implementation as a C++ library later~(\S\ref{subsec:impl_taint}).

\subsection{Data flow safety}
\label{subsec:dataflow}

All data originating from a sandbox begins life tainted.
Tainting is automatically applied by wrapping data with the \code{tainted<T>} constructor.
Tainting does not change the memory layout of a value, only its type.
Once applied, though, tainting cannot be removed.
The only way to remove a taint is through explicit
validation~(\S\ref{subsec:dataflow_validation}).
 
In general, \sys propagates taint following a standard IFC discipline.
For example, we propagate taint to any data derived from \tainted values such as data
accessed through a \tainted pointer, or arithmetic operations when one or more
operands are \tainted.
We detail how \sys implements \tainted types and taint tracking in the C++ type
system in Section~\ref{subsec:impl_taint}.
In the rest of this section we show how \sys uses tainting to ensure data flow
safety.

\para{Data flow into the renderer}
To protect the renderer from malicious inputs, all data flows from the sandbox
into the renderer are \tainted.
Data primarily flows out of the sandbox through two interfaces.
First, \code{sandbox_invoke()}, the only way to call into the sandbox, taints 
its return value.
Second, the use of \code{sandbox_callback()}, which permits callbacks into the
renderer from the sandbox, statically forces the parameters of callbacks to be 
\tainted.
Any code failing to follow either of these rules would cause a compilation 
error.

As an example, consider the JPEG decoder code that calls \jpeg's
\code{jpeg_read_header()} to parse headers shown below:
\begin{minted}[linenos=false, breaklines=true, mathescape=true,
escapeinside=||]{cpp}
jpeg_decompress_struct mInfo;
int status = jpeg_read_header(&mInfo, TRUE);
\end{minted}
\noindent
With \sys, the second line must be modified to use \code{sandbox_invoke()}, and
\code{status} must be declared as \tainted\footnote{
 In this paper, we use full type names such as
\footnotecode{tainted<int>} for clarity. In practice, we use C++'s
\footnotecode{auto} keyword to make code less verbose.}:
\begin{minted}[linenos=false, breaklines=true, mathescape=true,
escapeinside=||]{cpp}
tainted<int> status = sandbox_invoke(mRLBox, jpeg_read_header, &mInfo, TRUE);
\end{minted}

In addition to the invoke and callback interfaces, data can flow into the 
renderer
via pointers to sandboxed memory.
\sys, however, forces both these pointers and any data derived from them (e.g.,
via pointer arithmetic or pointer dereferencing) to be \tainted---and, as we
discuss shortly, using \tainted pointers in the renderer is always safe.

\para{Data flow into the sandbox}
\sys requires data flowing into sandbox from the renderer to either have a simple
numeric type or a \tainted type.
Untainted pointers, i.e., pointers into renderer memory, are not permitted.
This restriction enforces a code correctness requirement---sandboxed 
code only gets pointers it can access, i.e., pointers into sandbox
memory---and, moreover, preserves the renderer's ASLR: any accidental pointer
leaks are eliminated by construction.

Compile-time errors are used to guide the code changes necessary to use a
sandboxed library.
To demonstrate this, we continue with the example of JPEG header parsing shown
above. To start, note that the \code{TRUE} parameter to \code{jpeg_read_header()} can
remain unchanged as it has a simple numeric type (in C++). On the other hand,
the parameter \code{&mInfo} points to a \code{struct} in renderer memory, which
\jpeg cannot access; \sys thus raises a compile-time error.

To address this compilation error, \sys requires such shared data structures to
be allocated in sandbox memory using \code{sandbox_malloc()}:
\begin{minted}[linenos=false, breaklines=true, mathescape=true,
escapeinside=||]{cpp}
tainted<jpeg_decompress_struct*> p_mInfo = sandbox_malloc<jpeg_decompress_struct>(mRLBox);
tainted<int> status = sandbox_invoke(mRLBox, jpeg_read_header, p_mInfo, TRUE);
\end{minted}
\noindent
Placing shared data structures in sandboxed memory in this way 
simplifies data marshaling of pointer parameters during function 
calls---\sys simply marshals pointers as numeric types, it does not eagerly
copy objects.
Indeed, this design allows \sys to automatically generate the marshaling code
without any user annotations or pointer bounds information (as required by most
RPC-based sandboxing tools).
Moreover, \sys can do all of this without compromising the renderer's
safety---renderer code can only access shared sandbox memory via \tainted
pointers.

While \sys does not allow passing untainted pointers into \jpeg, 
pointers to callback functions, such as \code{fill_input_buffer()}, 
need to be shared with the sandbox---these can be shared either as 
function call parameters to \jpeg functions, return values from 
callbacks, or by directly writing to sandbox memory.
\sys permits this without exposing the raw callback pointers to \jpeg,
through a level of indirection: \emph{trampoline functions}.
Specifically, \sys automatically replaces each \ff callback passed to \jpeg with a pointer
to a trampoline function and tracks the mapping between the two.
When the trampoline function is invoked, \sys invokes the
appropriate \ff callback on \jpeg's behalf.

\para{Benefits of tainted pointers}
\label{subsec:dataflow_aslr}
By distinguishing pointers to renderer memory from pointers to sandbox memory
at the type level with \tainted, \sys can automatically enforce several
important security requirements and checks.
First, \sys does not permit \ff to pass untainted pointers to \jpeg.
Second, \sys automatically swizzles and unswizzles pointers appropriately when
pointers cross the renderer-library boundary, including pointers in deeply
nested data structures.
(We give a more detailed treatment of pointer swizzling in 
\appref{A}{sec:appendix_swizzling}.)
Third, \sys automatically applies pointer-bounds sanitization checks when
\tainted pointers are created to ensure they always point to sandboxed memory.
Together, these properties ensure that we preserve the renderer's ASLR---any
accidental pointer leaks are eliminated by construction---and that the renderer
cannot be compromised by unsanitized pointers---all \tainted pointers point to
sandbox memory.

\subsection{Data validation}
\label{subsec:dataflow_validation}
\ignore{
\begin{figure}
  \begin{tabularx}{\linewidth}{ p{2.7cm} | X}
    \toprule
    \textbf{Name \& Params} & \textbf{Description}\\
    \midrule
    \texttt{\small verify (verify\_fn)} &
    Validate already-copied primitive value.\\
    \hline
    \texttt{\small copyAndVerify (verify\_fn)} &
    Copy (single) primitive value and validate it.\\
    \hline
    \texttt{\small copyAndVerify (verify\_fn, nr)} &
    Copy \texttt{\small nr} elements from an array or buffer and validate.\\
    \hline
    \texttt{\small copyAndVerify (verify\_fn, len)} &
    Copy string (of length at most \texttt{\small len}) and validate.\\
    \hline
    \texttt{\small unsafeUnverified()} &
    Unsafely ignore taint wrapper.\\
    \bottomrule
  \end{tabularx}
  \caption{
    Unwrapping methods of \taintedW{T} values; the methods available
    depend on the type of value being unwrapped (primitive, pointer,
    struct, etc.).
  }
  \label{fig:unwrappingFuncs}
\end{figure}
}

\sys disallows computations (e.g., branching) on \tainted data that 
could affect the renderer control and data flow.
The \ff renderer, however, sometimes needs to use data produced by library code.
To this end, \sys allows developers to unwrap \tainted values, i.e., convert a
\taintedW{T} to an untainted \code{T}, using validation methods.
A validation method takes a closure (C++ lambda) that unwraps the tainted type 
by performing necessary safety checks and returning the untainted result. 
Unfortunately, it is still up to the user to get these checks correct; \sys 
just makes this task easier.

\sys simplifies the burden on the developer by offering different types of
validation functions.
The first, \code{verify(verify_fn)}, validates simple \tainted value types
that have already been copied to renderer memory (e.g., simple values),
as shown in this example:

\begin{minted}[linenos=false, breaklines=true, mathescape=true, escapeinside=~~]{cpp}
tainted<int> status = sandbox_invoke(mRLBox, jpeg_read_header, p_mInfo, TRUE);
int untaintedStatus = status.verify([](int val){
  if (val == JPEG_SUSPENDED ||
      val == JPEG_HEADER_TABLES_ONLY ||
      val == JPEG_HEADER_OK ) { return val; }
  else { /* DIE! */ }
});
if (untaintedStatus == JPEG_SUSPENDED) { ... }
\end{minted}
\noindent
Not all \tainted data lives in renderer memory, though. Validating shared
\tainted data structures that live in sandbox memory is unsafe: a concurrent
sandbox thread can modify data after it's checked and before it's used.
The \code{copyAndVerify(verify_fn, arg)} validator addresses this by copying
its arguments into renderer memory before invoking the \code{verify_fn}
closure.
To prevent subtle bugs where a \verify function is accidentally applied to data 
in shared memory, \sys issues a compile-time error---notifying the developer 
that \copyAndVerify is needed instead.

The \code{unsafeUnverified()} function removes tainting without any checks.
This obviously requires care, but has several legitimate uses.
For example, when migrating a codebase to use sandboxing, using \unsafe allows 
us to incrementally test our code before all validation closures have been
written (\S\ref{sec:migration}).
Furthermore, \unsafe is sometimes safe and necessary for performance---e.g., 
passing a buffer of \jpeg-decoded pixel data to the \ff renderer without 
copying it out of sandbox memory.
This is safe as pixel data is simple byte arrays that do not require complex 
decoding.

\para{Validation in the presence of double fetches}
Though our safe validation functions ensure that sandbox code cannot
concurrently alter the data being validated, in practice we must also account
for double fetch vulnerabilities.

Consider, for example, migrating the following snippet from the
\code{nsJPEGDecoder::OutputScanlines} function:
\begin{minted}[linenos=true, breaklines=false, mathescape=true,
escapeinside=||]{cpp}
while (mInfo.output_scanline < mInfo.output_height) {|\label{line:scanlinechck}|
  ...
  imageRow = reinterpret_cast<uint32_t*>(mImageData) +
    |$\hookrightarrow$|(mInfo.output_scanline * mInfo.output_width);|\label{line:scanlineuse}|
  ...
}
\end{minted}
Here, \code{mInfo} is a structure that lives in the sandbox shared memory.
Buffer \code{imageRow} is a pointer to a decoded pixel-row that \ff hands off to
the rendering pipeline and thus must not be \tainted.
To modify this code, we must validate the results on
lines~\ref{line:scanlinechck}~and~\ref{line:scanlineuse} which are tainted as
they rely on \code{mInfo}.
Unfortunately, validation is complicated by the double fetch: a concurrent
sandbox thread could change the value of \code{output_scanline} between its
check on line~\ref{line:scanlinechck} and its use on
line~\ref{line:scanlineuse}, for example.
Unsafely handling validation would allow the sandbox to control the value of
\code{imageRow} (the destination buffer) and thus perform arbitrary
out-of-bounds writes.

We could address this by copying \code{output_scanline} to a local
variable, validating it once, and using the validated value on both lines.
But, it's not always this easy---in our port of \ff we found numerous instances
of multiple reads, interspersed with writes, spread across different functions.
Using local variables quickly became intractable.

To address this, \sys provides a \code{freeze()} method on \tainted variables
and struct fields.
Internally, this method makes a copy of the value into renderer memory and ensures
that the original value (which lives in sandbox memory), when used, has not
changed.
To prevent accidental misuse of freezable variables, \sys disallows the
renderer from reading freezable variables and fields until they are
frozen.
\sys does, however, allow renderer code to write to frozen variables---an
operation that modifies the original value and its copy.
Finally, the \code{unfreeze()} method is used to restore the sandbox's write access.

Unlike most other \sys features, ensuring that a variable remains frozen
imposes some runtime overhead.
This is thus a compile-time, opt-in feature that is applied to
select variables and \code{struct} fields.


\para{Writing validators}
\label{subsec:validator_styles}
We identify two primary styles of writing validators in our porting \ff to use 
sandboxed libraries:
we can focus on either preserving application invariants or on preserving
library invariants when crossing the trust boundary.
We demonstrate these two alternate styles, using the above
\code{OutputScanlines} example.
\begin{CompactEnumerate}
\item \emph{Maintaining application invariants:}
The first focuses on invariants expected by \ff.
To do this, we observe that \code{imageRow} is a pointer into the
\code{mImageData} buffer and is used as a destination to write one row of pixels.
Thus, it is sufficient to ensure that the result of \code{output_scanline *
output_width} is between \code{0} and \code{mImageDataSize - rowSize}.
This means that the \code{imageRow} pointer has room for at least
one row of pixels.

\item \emph{Checking library invariants:}
The second option focuses on invariants provided by \jpeg.
This option assumes that the \ff decoder behaves correctly when \jpeg is
well-behaved.
Hence, we only need to ensure that \jpeg adheres to its specification.
In our example, the \jpeg specification states that \code{output_scanline} is
at most the height of the image: we thus only need to freeze
\code{output_scanline} and then validate it accordingly.
\end{CompactEnumerate}

\subsection{Control flow safety}
\label{subsec:controlflow}

As discussed in Section~\ref{sec:problems}, a malicious sandbox could attempt to
manipulate renderer control flow in several ways. While data attacks
on control flow are prevented by tainting (e.g., it's not possible to branch on a
\tainted variable), supporting callbacks requires additional support.

\para{Control transfers via callbacks}
It's unsafe to allow sandboxes to callback arbitrary functions in the renderer.
It's also important to ensure they can only call functions which use \tainted
appropriately. Thus, \sys forces application developers---via
compile-time errors---to explicitly register callbacks using the
\code{sandbox_callback()} function.
For instance, line~\ref{line:cd_register_cb} in Figure~\ref{fig:jpeg_decoder},
must be rewritten to:
\begin{minted}[linenos=false, breaklines=true, mathescape=true,
escapeinside=||]{cpp} 
mSourceMgr.fill_input_buffer = sandbox_callback(mRLBox,fill_input_buffer);
\end{minted}

Statically whitelisting
callbacks alone is insufficient---an attacker-controlled
sandbox could still corrupt or hijack the control flow of the renderer by
invoking callbacks at unexpected times.
To address this class of attacks, \sys supports unregistering callbacks with the
\code{unregister()} method.
Moreover, the framework provides RAII (resource acquisition is
initialization)~\cite{cpp_pl} semantics for callback registration, which allows
useful code patterns such as automatically unregistering callbacks after
completion of an invoked \jpeg function.

To deal with callback state exchange attacks~(\S\ref{sec:problems}), \sys
raises a compile-time error when renderer pointers leak into the sandbox.
For example, the JPEG decoder saves its instance pointer with \jpeg and retrieves
it in the \code{fill_input_buffer()} callback, as shown on
line~\ref{line:cd_client_data_decoder} of Figure~\ref{fig:jpeg_decoder}.
\sys requires the application developer to store such callback state
in the application's thread local storage (TLS) instead of passing it to \jpeg.
Thus, when \code{fill_input_buffer()} is invoked, it simply retrieves the
decoder instance from the TLS, preventing any pointer leaks or callback state
modifications.

\para{Non-local control transfers}
A final, but related concern is protecting control flow via
\code{setjmp()}/\code{longjmp()}. These functions are used for exception
handling (e.g., \code{my_error_exit()} in
Figure~\ref{fig:jpeg_decoder}). They work like a non-local \code{goto}, storing
various registers and CPU state in a \code{jmp_buf} on \code{setjmp()}
and restoring them on \code{longjmp()}.

Naively porting \jpeg and \png would store \code{jmp_buf} in sandboxed memory.
Unfortunately, this doesn't work---there is no easy way to validate a
\code{jmp_buf} that is portable across different platforms.
We thus instead place such sensitive state in the renderer's TLS and avoid
validation altogether.
With \jpeg this is straightforward since the \code{jmp_buf} is only used in the
\code{my_error_exit()} callback.
\png, however, calls \code{longjmp()} itself.
Since we can't expose \code{longjmp()} directly, when sandboxing \png, we
expose a \code{longjmp()} trampoline function that calls back to the renderer
and invokes \code{longjmp()} on \png's behalf, using the \code{jmp_buf} stored
in the TLS.

\section{Simplifying migration}
\label{sec:migration}

\sys simplifies migrating renderer code to use sandboxed libraries while 
enforcing appropriate security checks.
\sys does this by removing error-prone glue code (e.g., for data marshaling)
and by reducing the security-sensitive code required for migration.
The resulting reduction in developer effort is evaluated in detail in 
Section~\ref{subsec:migrate_effort}.

The \sys framework helps automate porting by (1) allowing \ff developers to {\it
incrementally} port application code---the entire application compiles and runs
with full functionality (passing all tests) between each step of porting---and (2)
guiding the porting effort with compiler errors which highlight what the next
step should be.
We illustrate how \sys minimizes engineering effort using this incremental,
compiler-directed approach using our running example. 

To start, we assume a standard \ff build that uses a statically linked \jpeg.

\para{Step 1 (creating the sandbox)} We start using \sys by creating a sandbox
for \jpeg using the \code{None} sandboxing architecture.  As the name suggests,
this sandbox does not provide isolation; instead it redirects all function
calls back to the statically linked, unsandboxed \jpeg.  However, \sys still
fully enforces all of its type-level guarantees such as tainting untrusted 
data. Thus, we can start using \sys while still passing functional tests.

\para{Step 2 (splitting data and control flow)} Next, we migrate each function 
call to \jpeg to use the \code{sandbox_invoke()} API.
\sys flags calls passing pointers to the sandbox as compile-time errors after 
this conversion, as the sandbox will be unable to access the (application) 
memory being pointed to.
To resolve this, we also convert the allocations of objects being passed to
\jpeg to instead use \code{sandbox_malloc()}.
For example, in Section~\ref{sec:problems}, we rewrote:
\begin{minted}[linenos=false, breaklines=true, mathescape=true,
escapeinside=||]{cpp}
tainted<int> status = sandbox_invoke(mRLBox, jpeg_read_header, &mInfo, TRUE);
\end{minted}
to, instead, allocate the \code{mInfo} object in the sandbox:
\begin{minted}[linenos=false, breaklines=true, mathescape=true,
escapeinside=||]{cpp}
tainted<jpeg_decompress_struct*> p_mInfo =
  sandbox_malloc<jpeg_decompress_struct>(mRLBox);
tainted<int> status = sandbox_invoke(mRLBox, jpeg_read_header, p_mInfo, TRUE);
\end{minted}

\noindent At this point, we need to re-factor the rest of this function and
several other JPEG decoder functions---\code{mInfo} is a data member of the
\code{nsJPEGDecoder} class.
Doing this in whole is exhausting and error-prone.
Instead, we remove the \code{mInfo} data member and add one extra line of code
in each member function before \code{mInfo} is first used:
\begin{minted}[linenos=false, breaklines=true, mathescape=true,
escapeinside=||]{cpp}
jpeg_decompress_struct& mInfo = *(p_mInfo.unsafeUnverified());
\end{minted}
This {\it unsafe alias pattern} allows the remainder of the function body to 
run 
unmodified, i.e., the alias defined in this pattern can be used everywhere the 
original \code{mInfo} variable is needed, albeit unsafely, as \unsafe 
temporarily suppresses the need for validation functions.

We also need to deal with return values from
\code{sandbox_invoke()} which are \tainted, either by writing
validation functions to remove the taint or deferring this till the next
step and using \unsafe to satisfy the type checker.  Again, the entire
application should now compile and run as normal.

\para{Step 3 (hardening the boundary)} Our next goal is to gradually remove all
instances of the unsafe alias pattern, moving \ff to a point where all data 
from the sandbox shared memory and all \tainted return values are handled safely.

We can do this incrementally, checking our work as we go by ensuring the
application still compiles and runs without errors.
To do this, we simply move each unsafe-alias pattern downwards in the 
function source; as it moves below a given statement, that statement is no longer
able to use the alias and must be converted to use the actual tainted value.
This may involve writing validation functions, registering callbacks, or 
nothing (e.g., for operations which are safe to perform on \tainted values).
We can compile and test the application after any or all such 
moves.
At the end, shared data is allocated appropriately, and all
\tainted values should be validated---no instances of \unsafe should remain.

\para{Step 4 (enabling enforcement)} Our final task is to replace the
\code{None} sandbox with one that enforces strong isolation.  To start, we
remove the statically-linked \jpeg and change the sandbox type from \code{None}
to \code{None_DynLib}.  In contrast to the \code{None} sandbox, the \code{None_DynLib}
sandbox dynamically loads \jpeg.  Any remaining calls to \jpeg made without the
\code{sandbox_invoke()} will fail with a symbol resolution error at compile
time. We resolve these and finish by changing the sandbox type to
\code{Process}, \code{NaCl} sandbox types that enforce isolation. 
We discuss these isolation mechanisms in more detail in Section~\ref{subsec:isol_mechanisms}.

\section{Implementation}
\label{sec:impl}

Our implementation consists of two components:
(1) a C++ library that exposes the APIs that developers use when sandboxing a
third-party library, and
(2) two isolation mechanisms that offer different scaling-performance
trade-offs.
We describe both of these below, and also describe a third approach in 
Section~\ref{sec:upstream}.

\subsection{\sys C++ API and type system}
\label{subsec:impl_taint}

The \sys API is implemented largely as a pure C++ library.
This library consists of functions like \code{sandbox_invoke()} that are used
to safely transfer control between the application and library.
These functions return \tainted values and can only be called with \tainted
values or primitives.
The library's wrapped types (e.g., \taintedW{T}) are used to ensure dataflow safety
(e.g., when using a value returned by \code{sandbox_invoke()}).
Since the implementation of the control flow functions is mostly standard, we
focus on our implementation of tainted values.

\para{The \taintedW{T} wrapper} 
We implement \taintedW{T} as a simple wrapper that preserves the memory layout
of the unwrapped \code{T} value, i.e., \code{tainted<int>} is essentially
\code{struct tainted<int> { int val; }}.
The only distinction between tainted and untainted values is at the type-level.
In particular, we define methods and operators on \taintedW{T} values that (1)
ensure that tainted values cannot be used where untainted values are expected
(e.g., branch conditions) without validation and (2) allow certain computations
on tainted data by ensuring their results are themselves tainted.

In general, we cannot prevent developers from deliberately abusing unsafe C++
constructs (e.g., \code{reinterpret_cast}) to circumvent our wrappers.
Our implementation, however, guards against common C++ design patterns that
could inadvertently break our \tainted abstraction.
For example, we represent tainted pointers as \taintedW{T*} and
not \mbox{\taintedW{T}\code{*}}.
This ensures that developers cannot write code that inadvertently unwraps 
\tainted pointers via pointer decay---since all C++ pointers can decay to 
\code{void*}.
We also use template meta-programming and SFINAE to express more complex
type-level policies.
For example, we disallow calls to \verify on pointer types and ensure that
callback functions have wrapped all parameters in \tainted.

\para{Operators on \tainted data}
For flexibility, we define several operators on \tainted types.
Operations which are always safe, such as the assignment (\code{operator=}) of
a \taintedW{int}, simply forward the operation to the wrapped \code{int}.
Other operators, however, require return types to be \tainted.
Still others require runtime checks.
We give a few illustrative examples.
\begin{CompactItemize}
  \item \emph{Wrapping returned values:}
  We allow arithmetic operators (e.g., \code{operator+}) on, say, \code{tainted<int>}s, or a
  \code{tainted<int>} and an untainted \code{int}, but wrap the return value.
  \item \emph{Runtime checks:}
  We allow array indexing with both \code{int}s and \code{tainted<int>}s by
  defining a custom array indexing operator \code{operator[]}. This operator
  performs a runtime check to ensure the array indexing is within sandbox
  memory.
  \item \emph{Pointer swizzling:}
  We also allow operations such as \code{operator=}, \code{operator*}, and \code{operator->} on tainted
  pointers, but ensure that the operators account for swizzling (in addition to
  the runtime bounds check) when performing these operations. As with the
  others, these operators return \tainted values.
  In \appref{A}{sec:appendix_swizzling}, we describe the subtle details
  of type-driven automatic pointer swizzling.

\end{CompactItemize}

\para{Wrapped structs}
Our library can automatically wrap primitive types, pointer types,
function pointers, and static array types.
It cannot, however, wrap arbitrary user-defined structs without some added
boilerplate definitions.
This is because C++ (as of C++17) does not yet support reflection on struct
field \emph{names}.
We thus built a \textasciitilde{}100LOC Clang plugin that automatically
generates a header file containing the required boilerplate for all struct
types defined in the application source.

\para{Other wrapped types}
In addition to the \tainted wrapper type, \sys also relies on several other
types for both safety and convenience.
As an example of safety, our framework distinguishes registered callbacks
from other function pointers, at the type level.
In particular, \code{sandbox_callback} returns values of type
\code{callback<...>}.
This allows us to ensure that functions that expect callbacks as arguments can
in fact only be called with callbacks that have been registered with
\code{sandbox_callback}.
As an example of convenience, \sys provides RAII types such as
\code{stack_arr<T>} and \code{heap_arr<T>} which minimize boilerplate.
With these types, developers can for instance invoke a function with an inline
string: \code{sandbox_invoke(sbox, png_error,} \code{stack_arr("..."))}.

\subsection{Efficient isolation mechanisms}
\label{subsec:isol_mechanisms}
The \sys API provides a plugin approach to support different, low-level
sandboxing mechanisms.
We describe two sandboxing mechanisms which allow portable and efficient
solutions for isolating libraries in this section.
In Section~\ref{sec:upstream} we describe a third mechanism, based on
WebAssembly, that we recently integrated in production \ff.
 
The first mechanism uses software-based fault isolation
(SFI)~\cite{wahbe-et-al:sfi:sosp93} extending Google's Native Client
(NaCl)~\cite{nacl,nacl-amd64}, while the second uses OS processes
with a combination of mutexes and spinlocks to achieve performance.
These approaches and trade-offs are described in detail below.

\para{SFI using NaCl}
\label{sec:impl_sfi}
SFI uses inline dynamic checks to restrict the memory addressable by a library 
to a subset of the address space, in effect isolating a library from 
the rest of an application within a single process.
SFI scales to many sandboxes, has predictable performance, and incurs
low overhead for context switching (as isolation occurs within a single process).
The low context-switch overhead (about 10$\times$ a normal function call) is
critical for the performance of streaming libraries such as \jpeg, which tend to
have frequent control transfers.
We explore this in more detail later (\S\ref{sec:eval}). 

To support library sandboxing with SFI, we extend the NaCl compiler
toolchain~\cite{nacl,nacl-amd64}.
NaCl was originally designed for sandboxing mobile
code in the browser, not library sandboxing.
Hence, we made significant changes to the compiler, optimization passes, ELF
loader, machine model and runtime; we give a thorough description of these changes
in \appref{B}{sec:appendix_nacl}.
To ensure that our changes do not affect security, we always verify the code
produced by our toolchain with the \emph{unmodified} NaCl binary code verifier.

We use our modified NaCl compiler toolchain to compile libraries like \jpeg along 
with a custom runtime component.
This runtime component provides symbol resolution and facilitates
communication with the renderer.

\para{Process sandboxing}
\label{sec:impl_spi}
Process sandboxing works by isolating a library in a separate \emph{sandbox 
process} whose access to the system call interface is restricted using 
seccomp-bpf~\cite{mozilla-sandbox}.
We use shared memory between the two processes to pass function arguments and
to allocate shared objects.
Compared to SFI, process sandboxing is simpler and does not need custom compiler
toolchains.
When used carefully, it can even provide performance comparable to SFI
(\S\ref{sec:eval}).

As with SFI, process sandboxing also includes a custom runtime that handles
communication between the library and renderer.
Unlike SFI, though, this communication is a control transfer that
requires inter-process synchronization.
Unfortunately, using a standard synchronization mechanism---notably, condition
variables---is not practical: a simple cross-process function is over
300$\times$ slower than a normal function call. 

Our process sandbox uses both spinlocks and condition variables, allowing users
to switch between to address application needs.
Spinlocks offer low control-transfer latency (20$\times$ a normal function 
call), at the cost of contention and thus scalability.
Condition variables have higher latency (over 300$\times$ a normal function
call), but minimize contention and are thus more scalable.
In the next section we detail our \ff integration and describe how we
switch between these two process sandboxing modes.

\subsection{Integrating \sys with \ff}
\label{subsec:integration}
To use the SFI or Process sandbox mechanisms efficiently in 
\ff, we must make several policy decisions about when to create and destroy 
sandboxes, how many sandboxes to keep alive, and for the Process sandbox when 
to switch synchronization modes.
We describe these below.

\para{Creating and destroying sandboxes}
We build \ff with \sys sandboxing web page decompression, image decoding, and
audio and video playback.
We apply a simple policy of creating a sandbox on demand---a fresh sandbox is 
required when decoding a resource with a unique \code{<renderer, library, 
content-origin, content-type>} as discussed in Section~\ref{sec:fine_grain}.
We lazily destroy unused sandboxes once we exceed a fixed threshold.
We determine this threshold experimentally.
Most webpages have a large number of compressed media files as compared to the
number of images (\S\ref{subsec:cross_origin}).
Since we can only reasonably scale to 250 sandboxes (\S\ref{subsubsec:scaling}),
we conservatively use a threshold 10 sandboxes for JPEG and PNG image decoding,
and 50 sandboxes for webpage decompression.
Browsers do not typically play multiple audio or video content
simultaneously---we thus simply create a fresh sandbox for each audio and video
file that must be decoded and destroy the sandbox immediately after.

\para{Switching synchronization modes}
For the Process sandbox, we switch between 
spinlocks and conditional variables according to two policies.
First, we use spinlocks when the renderer performs a latency sensitive 
task, such as decoding an image using a series of synchronous \jpeg function
calls and callbacks.
But, when the renderer requests more input data, we switch to the condition
variables; spinlocks would create needless CPU contention while waiting for
this data (often from the network).
Second, for large media decoding such as 4K images, we use condition variables:
each function call to the sandbox process takes a large amount of time
(relative to the context switch) to perform part of the decoding.
Though we can use more complex policies (e.g., that take load into effect), we
find these two simple policies to perform relatively well (\S\ref{sec:eval}).

\para{Leveraging multiple cores}
Since 95\% of devices that run \ff have more than 1 core, we can use multiple
cores to optimize our sandbox performance.\footnote{See
\url{https://data.firefox.com/dashboard/hardware}, last visited May 15, 2019.}
In particular, our process sandbox implementation pins the sandboxed process on
a separate CPU core from the renderer process to avoid unnecessary context
switches between the renderer and sandboxed process.
This is particularly important when using spinlocks since the sandbox process's
spinlock takes CPU cycles even when ``not in use''; pinning this process to a
separate core ensures that the sandbox process does not degrade the renderer
performance.
In our evaluation (\S\ref{sec:eval}), we reserve one of the system's cores
exclusively for the processes created by our sandboxing mechanism, when
comparing against the stock and SFI builds (which use all cores for the
renderer process' threads).

\section{Evaluation}
\label{sec:eval}

We present the following results:
\begin{CompactItemize}
\item {\it Cross origin resources that could compromise the renderer are
    pervasive}, and could even be used to compromise websites like Gmail.  We
     present measurements of their prevalence in the Alexa top 500 websites
     in section~\ref{subsec:cross_origin}.

\item {\it Library sandboxing overheads are modest}:
    section~\ref{subsec:individual} breaks down the individual sources of
	sandboxing overhead, section~\ref{subsec:end2end}, shows end-to-end page
	latencies and memory overheads for popular websites are modest with
	both isolation mechanisms(\S\ref{subsec:isol_mechanisms})--even on media
	heavy websites, section~\ref{subsec:micro}, shows that
	the CPU overhead of web page decompression and image decoding in the
	renderer process are modest, and that CPU and memory overheads scale
	well up to our current maximum of 250 concurrent sandboxes.

\item {\it Migrating a library into \sys typically takes a few days with modest
effort}, as shown in Section~\ref{subsec:migrate_effort}.

\item {\it \sys is broadly useful for library sandboxing beyond
\ff}, as demonstrated in ~\ref{subsec:non_firefox_sbx}, where we discuss our
experience apply \sys to sandboxing native libraries in Apache and
Node.js modules.

\end{CompactItemize}

\para{Machine Setup}
All benchmarks run on an Intel i7-6700K (4~GHz) machine with 
64 GB of RAM, and hyperthreading disabled, running 64-bit Ubuntu 
18.04.1.
\ff builds run pinned on two isolated CPU cores (i.e.,
no other process is allowed to run on these CPUs), to reduce noise.
As discussed in Section~\ref{subsec:integration}, with the process sandbox build,  
the renderer process is pinned to one core, while the sandbox process uses 
the other core.

\subsection{Cross-origin content inclusion}
\label{subsec:cross_origin}
\begin{figure}
\includegraphics[width=1\columnwidth]{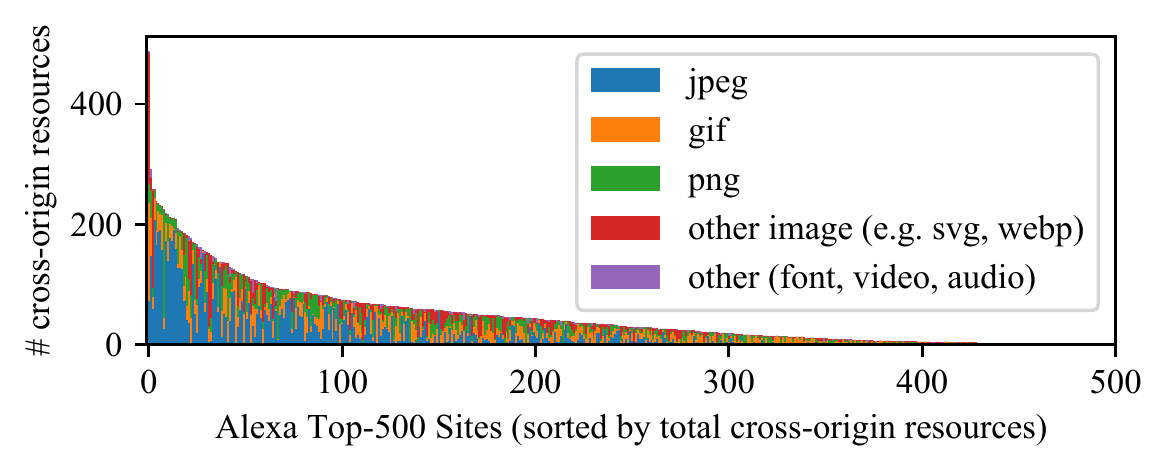}
\caption{Cross-origin resource inclusion in the Alexa Top-500.
  }
\label{fig:cross_origin}
\end{figure}
To evaluate how often web sites include content (e.g., images, audio, and
videos) cross-origin, we crawled the Alexa top 500 websites.
Our crawler---a simple \ff extension---logs all cross-origin resource requests
made by the website for 10 seconds after page load (allowing some dynamic
content to be loaded).
Figure~\ref{fig:cross_origin} shows our measurements, categorized by the
resource MIME type.

We find that 93\% of the sites load at least one cross-origin media resource (primarily images),
with mean of 48 and median of 30, cross-origin media
resources loaded.  Many of the resource loads (median 35 and mean 17) are not just cross-origin but also \emph{cross-site}.
In the presence of media parsing library bugs, such loads would undermine Site 
Isolation protections.

The pervasive use of cross-origin resources inclusion indicates that sandboxing
libraries at the \code{<renderer, library, content-origin, content-type>} granularity
can significantly reduce the renderer attack surface.
Although not all cross-origin content is necessarily untrusted, the origin is
nevertheless an important trust boundary in practice---and many websites do
consider cross-origin media untrusted.
For instance, Google allows users to freely upload content to
\code{sites.google.com}, but serves such media content from
\code{googleusercontent.com}.
Google even re-encodes images, another sign that such images are untrusted.
 
Unfortunately, they do not re-encode video or audio files.
To test this, we created a page on \code{sites.google.com} in which we 
embedded 
both the VPX video proof of concept exploit of CVE-2015-4506~\cite{vpxbug}
and the OGG audio proof of concept exploit of CVE-2018-5148~\cite{oggbug}.
In both cases the files were unmodified.
For VPX, we modified \ff and Chrome (with Site Isolation) to 
re-introduce the 
VPX bug and visited our malicious website: in both browsers the video 
successfully triggered the bug.
 
We found we could include such malicious content as part of an
email to a Gmail address.
Gmail re-encodes images, but does not re-encode the video and audio files.
The Gmail preview feature even allows us to play the audio track---which 
surprisingly, we found was hosted on \code{mail.google.com}.

\subsection{Baseline \sys overhead}
\label{subsec:individual}

To understand the overhead of different parts of the \sys framework in
isolation we perform several micro-benchmarks.

\para{Sandbox creation overhead} The overhead of sandbox creation is
$\approx$1ms for SFI sandboxes and $\approx$2ms for process sandboxes.  These
overheads do not affect page latency for any realistic website, and can be
hidden by pre-allocating sandboxes in a pool. For this reason, we don't include
sandbox creation time in the remaining measurements.

\para{Control transfer overhead}
To understand the overhead of a control transfer, we measure the elapsed time
for an empty function call with different isolation mechanisms. For reference,
an empty function call without sandboxing takes about 0.02$\mu$s in our setup.
With sandboxing, we measured this to be 0.22$\mu$s for SFI sandboxes, 
0.47$\mu$s for Process sandboxes using spinlocks, and 7.4$\mu$s for Process 
sandboxes using conditional variables.
SFI and Process sandboxes using spinlocks are an order of magnitude slower than 
a normal function call, but over 10$\times$ faster than Processes using 
condition variables, and are thus better suited for workloads with frequent 
control transfers.

\para{Overhead of \sys dynamic checks} The \sys API introduces small overheads with its
safety checks (e.g., pointer bounds checks, swizzling, and data validation
functions).
To measure these, we compare the overhead of rendering \code{.jpeg} and
\code{.png} images on \ff with sandboxed libraries with and without \sys
enabled.
We find the difference to be negligible ($<1\%$).  This is unsurprising:
most of \sys's checks are static and our dynamic checks are
lightweight masks.

\para{Overhead of SFI dynamic checks}
Unlike process-based sandboxing, SFI incurs a baseline overhead (e.g., due to
inserted dynamic checks, padding etc.)~\cite{nacl}.  To understand the
overhead of our NaCl SFI implementation, we implement a small \code{.jpeg}
image decoding program and measure its slowdown when using a sandboxed \jpeg.
We find the overhead to be roughly 22\%.

\subsection{Migrating \ff to use \sys}
\label{subsec:migrate_effort}
\begin{figure*}
\footnotesize

\resizebox{\linewidth}{!}{%
\begin{tabular} {l|p{4.9cm}|c|c|c|c|c|c}
\toprule
& \textbf{\pbox{\textwidth}{Task}}
& \textbf{\pbox{\textwidth}{JPEG\\Decoder}}
& \textbf{\pbox{\textwidth}{PNG\\Decoder}}
& \textbf{\pbox{\textwidth}{GZIP\\Decompress}}
& \textbf{\pbox{\textwidth}{Theora\\Decoder}}
& \textbf{\pbox{\textwidth}{VPX\\Decoder}}
& \textbf{\pbox{\textwidth}{OGG-Vorbis\\Decoder}}
\\

\midrule
\multirow{5}{*}{\textbf{\pbox{\textwidth}{Effort saved by\\\sys automation}}}

&Generated marshaling code
& 133 LOC & 278 LOC & 38 LOC & 39 LOC & 60 LOC & 59 LOC \\
&Automatic pointer swizzles for function calls
& 30   & 96   & 5   & 36  & 46  & 34 \\
&Automatic nested pointer swizzles
& 17   & 5    & 6   & 8   & 9   & 5 \\
&Automatic pointer bounds checks 
& 64 checks & 25 checks & 8 checks & 12 checks & 15 checks & 14 checks\\
&Number of validator sites found
& 28   & 51   & 10  &  5  & 2   & 4 \\

\midrule
\multirow{4}{*}{\textbf{\pbox{\textwidth}{Manual effort}}}

&Number of person-days porting to \sys
& --   & --   & 1 day & 3 days & 3 days & 2 days \\
&Application LOC before/after port
& 720 / 1058 & 847 / 1317 & 649 / 757 & 220 / 297 & 286 / 368 & 328 / 395 \\
&Number of unique validators needed
& 11   & 14   & 3   & 3 & 2 & 2 \\
&Average LOC of validators
& 3 LOC & 4 LOC & 2 LOC & 3 LOC & 2 LOC & 2 LOC \\

\bottomrule 

\end{tabular}
  }

\caption{
  Manual effort required to retrofit \ff with fine grain isolation,
  including the effort saved by \sys's automation.
  %
  %
  %
  We do not report the number of days it took to port the JPEG and PNG decoders
  since we ported them in sync with building \sys.
}

\label{tab:effort}
\end{figure*}

\begin{figure*}[t]
  \begin{subfigure}{0.49\textwidth}
    \includegraphics[scale=0.95]{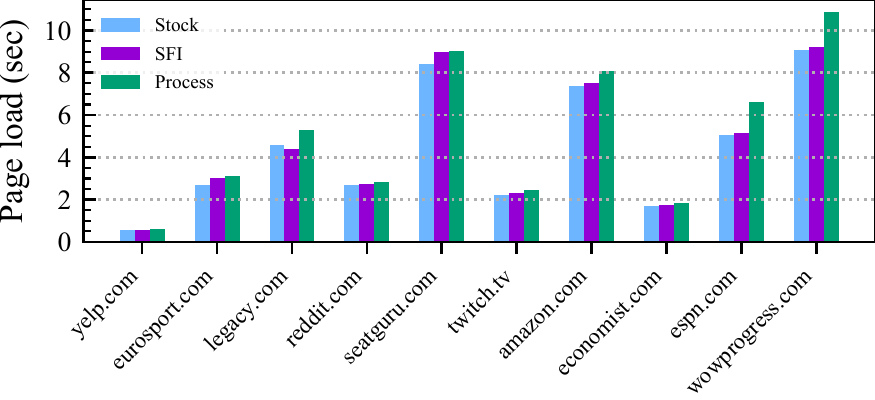}
    \label{fig:page_latency}
  \end{subfigure}
  \begin{subfigure}{0.49\textwidth}
    \includegraphics[scale=0.95]{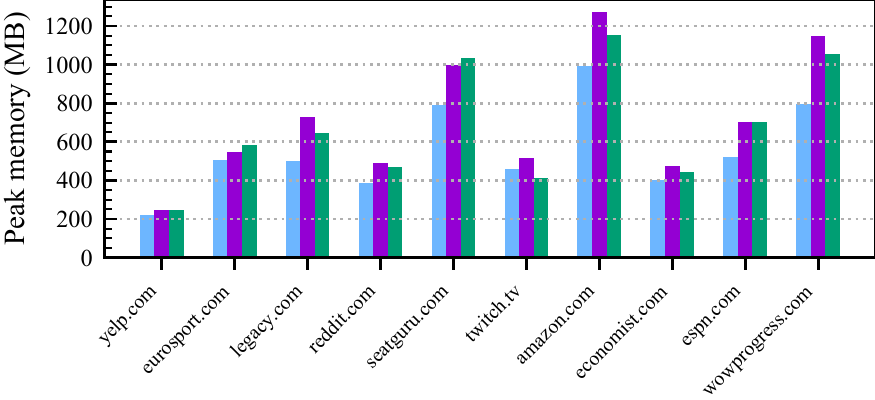}
    \label{fig:page_memory_overhead}
  \end{subfigure}
  \setlength{\abovecaptionskip}{-1pt}
  \caption{Impact of sandboxing on page load latencies and \emph{peak} memory 
  usage
  overheads. \ff with SFI sandboxes incurs a 3\% page latency and 25\% memory
  overhead while \ff with process isolation incurs a 13\% page latency and a
  18\% memory overhead.
  }
  \label{fig:page_overhead}
\end{figure*}  
In this section, we evaluate the developer effort required to migrate \ff to
use sandboxed libraries.
In particular, we report the manual effort required by \sys developers and the
manual effort saved by using the \sys API.
Figure~\ref{tab:effort} gives a breakdown of the effort.
On average, we find that it takes a bit over two days to sandbox a library with
\sys and roughly 180~LOC (25\% code increase); much of this effort is
mechanical (following Section~\ref{sec:migration}).
 
Figure~\ref{tab:effort} also shows the tasks that \sys automates away.
First, \sys eliminates 607 lines of glue-code needed by both the
Process and SFI sandboxes to marshal function parameters and return values for
each cross-boundary function call; \sys automatically generates this
boilerplate through C++ templates and meta-programming.
Second, \sys automatically swizzles pointers.
This is necessary for any sandbox functions or callbacks that accept pointers;
it's also necessary when handling data-structures with nested pointers that are
shared between the application and the sandbox.
This is a particularly challenging task without \sys, as manually identifying
the 297 locations where the application interacts with such pointers would have
been tedious and error-prone.
Third, \sys automatically performs bounds checks (\S\ref{subsec:dataflow});
the number of required pointer bounds checks that were automatically performed
by \sys are again in the hundreds (138).
 
Finally, \sys identifies the (100) sites where we must validate tainted data
(\S\ref{subsec:dataflow_validation}).
Though \sys cannot automate the validators, we find that we only need 35 unique
validators---all less than 4 lines of code.
In practice, we found this to be the hardest part of migration since it
requires understanding the domain-specific invariants.

\subsection{\sys overhead in \ff}
\label{subsec:end2end}

We report the end-to-end overheads of \ff with library sandboxing by 
measuring page latencies of webpages, memory overheads in \ff as well 
as audio video playback rates.

\para{Experimental setup}
We evaluate end-to-end performance with six sandboxed libraries: \turbo 
1.4.3, \libpng 1.6.3, \zlib 1.2.11, \vpx 1.6.1, \theora 1.2, and 
\vorbis as used in \ff 57.0.4.
We report performance for two \ff builds that use libraries sandboxed 
with SFI and Process mechanisms respectively.
Both builds create fresh sandboxes for each \code{<renderer, library, 
origin, content-type>} combination as described in 
\S\ref{sec:fine_grain}.
We measure impact on page load times for both these builds.

\subsubsection{End-to-end impact on real-world websites}
\label{subsubsec:realsites}
\para{Benchmark}
We report the overhead of page load latency and memory overheads in \ff 
with the six sandboxed libraries by measuring latencies of the 11 
websites used to measure the overhead of Site 
Isolation~\cite{site-isolation-usenix}.
These websites are a representative sample of both popular and
slightly-less-popular websites, and many of them make heavy use of media
resources.
We measure page load latency using \ff's Talos test 
harness~\cite{firefox-talos}.
We measure memory overheads with \code{cgmemtime}~\cite{cgmemtime}---in
particular, the peak resident memory and cache usage during a run of \ff.
We run the test 10 times and report the median values of page latency and
memory overheads.

\para{Results}
As shown in Figure~\ref{fig:page_overhead}, the page latency and CPU
utilization overheads are modest.
Our SFI build incurs a 3\% overhead in page latency while the Process sandbox
incurs an overhead of 13\%.
As a comparison point, the overhead of naively using process sandboxes (using
only conditional variables without any CPU pinning) incurs an overhead of
167\%.

We find the average peak renderer memory overhead to be $25\%$ and $18\%$
for the SFI and Process sandboxes, respectively.
This overhead is modest and, more importantly, transient: we can destroy a
sandbox after the media has been decoded during page load.

\subsubsection{Sandboxing video and audio decoding}
\label{subsubset:av}
To understand the performance overhead of \sys on video and audio decoding,
we measure the performance of \ff when decoding media from two video formats
(Theora and VPX) and one audio format (OGG-Vorbis).

\para{Benchmark} 
Our benchmark measures decoding performance on the Sintel sequence in 
Xiph.org's media test suite\footnote{Online: \url{https://media.xiph.org/}. 
Last visited November 15, 2019.}.
As this sequence is saved as lossless video, we setup the benchmark by 
first 
converting this to ultra HD videos of 4K resolution in the Theora and 
VP9 
formats; we similarly convert the lossless audio to a high resolution 
OGG-vorbis audio file with a sample rate of 96 kHz and a bit rate of 
500 Kb/s 
using the suggested settings for these formats~\cite{theora-encoding, 
vp9-encoding}.
Our benchmark then measures the frame-rate and bit-rate of the video 
and audio
playback---by instrumenting the decoders with timers---when using the sandboxed
libraries.
We run the test 5 times and report the median values of frame-rate and 
bit-rate.

\para{Results}
We find that neither the SFI nor the Process sandboxing mechanism visibly 
degrades 
performance. In particular, our sandboxed \ff are able to maintain the 
same 
frame-rate (24 fps for the VPX video and 60 fps for the Theora video) and 
bit-rate (478 bits per second) as \unmodff for 
these media files.

\subsection{Microbenchmarks of RLBox in \ff}
\label{subsec:micro}

\begin{figure*}[t]
\begin{subfigure}{0.49\textwidth}
  \includegraphics{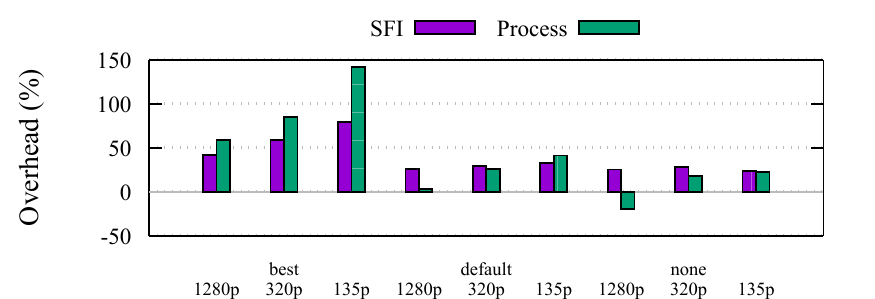}
  \caption{JPEG rendering overhead} \label{fig:jpeg_perf}
\end{subfigure}
\begin{subfigure}{0.49\textwidth}
  \includegraphics{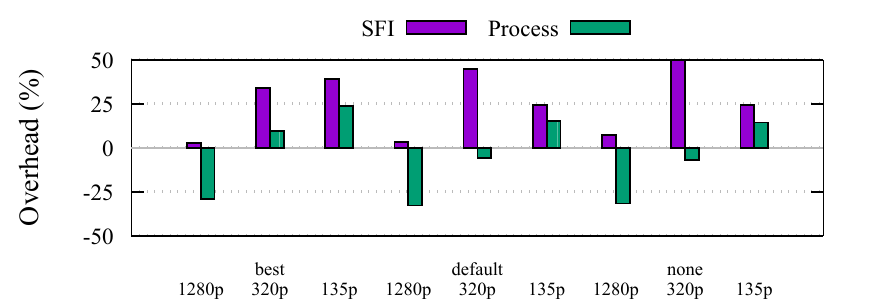}
  \caption{PNG rendering overhead} \label{fig:png_perf}
\end{subfigure}
\caption{Per-image decoding overhead for images at 3 compression levels 
and 3 resolutions, normalized against stock \ff.
\ifExtended
For most JPEGs, the SFI and Process sandbox overhead is 23-32\%, and 
$<$~41\% 
respectively.
For PNGs, the SFI and Process sandbox overhead is 2-49\% and $<$~15\% 
respectively.
We also observe some negative overheads and pathological cases where overhead
is as high as 140\%.
The reasons for this are explained in Section~\ref{subsubsec:images}. 
\fi
}
\label{fig:jpeg_png_perf}
\end{figure*}

To understand the performance impact of \sys on the different libraries, we
perform several microbenchmarks that specifically measure the impact of 
sandboxing webpage decompression, image decoding and sandbox scaling in \ff.

\subsubsection{Sandboxing webpage decompression}
\label{subsubsec:compression}

\ff uses \zlib to decompress webpages.
Since webpage decompression is done entirely before the page is rendered, we
report the overhead of sandboxing \zlib by measuring the slowdown in page
load time.

\para{Benchmark}
We create a webpage whose HTML content (excluding media and scripts) is 
1.8~MB, 
the size of an average web page\footnote{See the HTTP Archive page 
weight 
report, \url{https://httparchive.org/reports/page-weight}.  Last 
visited May 
15, 2019.}, and measure the page load time with Talos.
We use the median page load time from 1000 runs of this test. 

\para{Results} For both SFI and Process sandboxing mechanisms, the overhead of 
sandboxing \zlib is under 1\%.
In other words, the overhead of sandboxing \zlib is largely offset by other
computations needed to render a page.

\subsubsection{Sandboxing image decoding}
\label{subsubsec:images}

To understand performance impact of sandboxing on image rendering, we measure
per-image execution time for the \code{.jpeg} and \code{.png} decoders, with
different forms of sandboxing, and compare our results to \unmodff.
Decoder execution time is a better metric for image rendering performance than
page load time because \ff decodes and renders images asynchronously; the usual
test harness would notify us that a page has been loaded before images have
actually been decoded and displayed in full---and this might be visible to the user.

\para{Benchmarks}
We use the open Image Compression benchmark suite\footnote{Online:
\url{https://imagecompression.info/test_images/}.  Last visited
May~15, 2019.}  to measure sandboxed image decoding overheads.
To capture the full range of possibilities for performance overhead,
we measure overheads of images at 3 sizes (135p, 320p, and 1280p) and
three compression levels (best, default, and none) for each image in
the benchmark suite.
We run this test 4000 times for each image and compare the median 
decoder code 
execution times.

\para{Results}
Since all images in the suite produce similar results, we give the results of
one 8-bit image in Figure~\ref{fig:jpeg_png_perf}.
We start with three high-level observations.
First, both SFI and Process based sandboxes have reasonable overheads---23-32\% 
and $<$~41\% respectively for most JPEGs, 2-49\% and $<$~15\% 
respectively for PNGs.
Second, Process sandbox sometimes has negative overheads.
This is because the Process sandbox dedicates one of the two available cores
exclusively for execution of sandboxed code (\S\ref{subsec:integration})
including the sandboxed image rendering code, 
while the stock and SFI \ff builds use all cores evenly.
Third, for JPEGs at the best compression, the overhead relative to stock \ff 
is high---roughly 80\% for SFI and 140\% for Process sandboxes.
This is because decoding high compression images have low absolute decode 
times (\textasciitilde650$\mu$s), and thus have larger overheads as control 
transfer overheads between \ff and the sandbox image libraries cost are not 
effectively amortized.
However, in absolute terms, the differences are less than 1.5ms and have no 
impact on end-user experience.

\begin{figure}
    \includegraphics[scale=1]{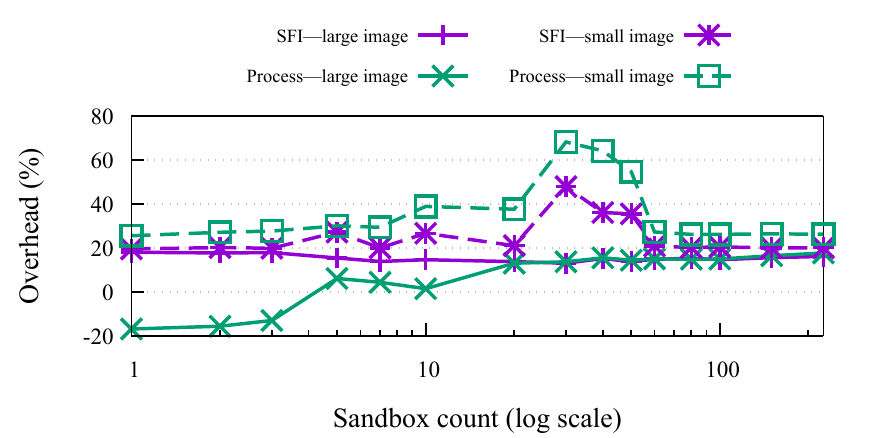}
  \caption{
    Performance overhead of image decoding with increasing the number of
    sandboxes (each image is rendered in a fresh sandbox).
  }
  \label{fig:sandbox_scaling}
\end{figure}

\subsubsection{Sandbox scaling characteristics}
\label{subsubsec:scaling}
Web pages often contain tens of images of different types from
multiple origins. Thus, the scaling properties of
different isolation mechanisms are an important consideration.

\para{Benchmark}
We evaluate sandbox scaling by rendering pages with an increasing number of
JPEG images from unique origins.
Each image thus creates a new sandbox which incurs both CPU and memory costs.
CPU costs are measured by measuring the total amount of time executing image 
decoding functions.
We measure memory overhead as before, but don't destroy any sandbox; this
allows us to estimate the worst case scenario where memory usage is \emph{not}
transient.
As before (\S\ref{subsubsec:images}), we measure the decoder execution time for
4000 image loads at each scale, and report the median overhead.

\para{Results}
Figure~\ref{fig:sandbox_scaling} shows the CPU overhead of image rendering as we
increase the number of sandboxes for both large~(1280p) and small~(135p) JPEG 
images using default compression.
This experiment allows us to make several observations.
We can run up to 250 concurrent SFI sandboxes before we run into limitations
like exhausting pre-allocated thread local storage or finding aligned free
virtual memory.
These limitations can be overcome with more engineering effort.
We never came close to these limits browsing real websites, including those of
Section~\ref{subsubsec:realsites}.
Both the SFI and the Process sandbox similarly scale well on both small and large 
images, with CPU overheads between 20\% and 40\% for most sandbox counts.
The process sandbox, however, scales only because we use multiple
synchronization modes described (\S\ref{sec:impl}).

Extra sandboxes add memory overhead for two reasons. First, each sandbox uses a
private copy of code (e.g., \jpeg and libc for each \jpeg sandbox). Second,
each sandbox has its own stack and heap.
In this experiment, we observed that memory consumption increases
linearly with the number of images (which corresponds to the number of 
sandboxes created).
On average, an SFI sandbox consumes 1.6 MB, while Process sandboxing consumes
2.4 MB for each sandbox.
Several optimizations to reduce memory consumption exist that we have not yet
implemented.
For example, the SFI sandbox currently loads a fresh copy of the code for each 
sandbox instance.
We could optimize this by sharing code pages between sandboxes---and, indeed,
we do this in production for our Wasm sandbox.

\subsection{\sys outside \ff}
\label{subsec:non_firefox_sbx}
\sys is a
general-purpose sandboxing framework that can be used in any C++ application.
To demonstrate its utility beyond \ff, we applied it in two different contexts: the Apache
web server and Node.js runtime.

Apache allows developers to write C modules that 
extend its 
base functionality.
These modules often depend on third-party libraries.
For example, the \modmarkdown~\cite{mod_markdown} module uses the \libmarkdown 
library to transform Markdown to HTML on the fly to support serving
Markdown files.

To protect Apache from bugs in \libmarkdown we modify
\modmarkdown to run \libmarkdown in an \sys SFI sandbox. The
change required a single person-day, and added or modified roughly 300 lines
of code.
We measured the average and tail latency as well as throughput of the webserver
using the autocannon 4.4.0 benchmarking tool~\cite{autocannon} with default 
configurations (1 minute runtime, 10 parallel connections) serving a 16K 
markdown file.
The unmodified webserver's average latency, tail latency and throughput were
10.5ms, 36ms and 940 requests/second, respectively; the sandboxed server's
average latency, tail latency and throughput were 14ms, 40ms and 684
requests/second.
Though the average latency and throughput overhead is modest, we observe that
the tail latency---arguably the most important metric---is within 10\% of baseline.

Node.js is a JavaScript runtime system written in C++, largely used for web 
applications.
Like Apache, Node.js allows developers to expose new functionality implemented 
in native plugins to the JavaScript code.
For example, the \textsf{bcrypt}~\cite{node_bcrypt} password hashing library 
relies on native code---indeed the JavaScript code largely wraps Provos'
C \textsf{bcrypt} library.
To protect the runtime from memory-safety bugs in the C library, we modify the 
C++ code in \textsf{bcrypt} to run the \textsf{bcrypt} C library in an \sys SFI
sandbox---a change that required roughly 2 person hours, adding or modifying 75
lines of code.
We measured---using the benchmark.js library---the overhead in average hashing 
throughput (hashing a random 32-byte password) to be modest: 27\%.

\section{Related work}
\label{sec:related}
\para{Isolation in the browser}
Modern browsers since Chrome~\cite{barth-et-al:chromium:08} rely on 
coarse grain
{\it privilege separation}~\cite{provos_openssh} to prevent browser compromises
from impacting the local OS~\cite{reis-isolating}.  However, a compromised
renderer process can still use any credentials the browser has for
other sites, enabling extremely powerful \emph{universal cross-site scripting}
(UXSS) attacks~\cite{UXSS}.

In response to UXSS attacks and recent Spectre attacks, Chrome introduced
\emph{Site Isolation}~\cite{site-isolation-usenix}.  Site Isolation puts pages
and iframes of different sites into separate processes. Unfortunately, as
discussed in Section~\ref{sec:intro}, this does not prevent UXSS attacks across
related sites (e.g., {\tt mail.google.com} and {\tt pay.google.com}).
Firefox's Project Fission~\cite{site-isolation-firefox} proposes to go
further and isolate at the origin boundary, similar to previous research
browsers~\cite{OP-browser, gazelle, ibos}, but this still does not protect the
renderer when loading cross-origin resources such as images.

An unpublished prototype using SFI called MinSFI~\cite{minsfi} was developed at
Google in 2013 to protect the Chrome renderer from compromise of \zlib
library; however, it was missing several features necessary for compatibility
and efficiency, including threading and callback support.
Additionally, the project was primarily focused on improving the efficiency of 
SFI rather than the integration challenges tackled by \sys including handling 
\tainted data, migration of code bases, etc.

In some parts of the renderer, there is no substitute for strong memory safety
to prevent attacks. Servo~\cite{servo-rust} is an ongoing project to
rewrite much
of the \ff rendering stack in Rust. However, for the foreseeable future, \ff and
other browsers will continue to rely on libraries written in C/++.
This makes sandboxing the most viable approach to containing
vulnerabilities.

\para{Sandboxing}
There has been some related work on providing APIs to simplify using sandboxed
libraries (e.g., Codejail~\cite{codejail} and Google Sandboxing
APIs~\cite{google-sandboxed-api}). However, these efforts do not provide the
type-driven automation of \sys (e.g., pointer swizzling and migration
assistance) nor the safety of \tainted types---leaving developers to manually
deal with attacks of Section~\ref{sec:problems}.
Sammler et al.~\cite{deepak-popl20} formal model addresses some of these
attacks using a type-direct approach, but require applications to be formally
verified correct (in contrast to our local validators) to give meaningful
guarantees.

There is a long line of work on sandboxing mechanisms with different
performance trade-offs~\cite{wahbe-et-al:sfi:sosp93, nacl-amd64, minsfi, wasm,
erim}.
Recent, excellent surveys~\cite{tan-sfi-survey, shu-isolation-survey} present a
comprehensive overview of these mechanisms.
\sys makes it easy for developers to use such mechanisms without modifying the
application or library (\S\ref{subsec:isol_mechanisms}).
In production we use WebAssembly; WebAssembly stands out as a principled
approach with wide adoption~\cite{wasm}.

\para{Data sanitization and leaks} There is a large body of work on
static and dynamic approaches to preventing or mitigating missed
sanitization errors; see the survey by Song
et~al.~\cite{sanitization-survey}. These tools are orthogonal to \sys.
Developers could use them to check their data
validation functions for bugs.

DUI Detector~\cite{DUI} uses runtime trace analysis to identify missing pointer
sanitizations in user code. Other work has looked at sanitizing user pointers
in the kernel~\cite{aiken-kernel-ptr}. For example, type
annotations~\cite{cqual-kernel-ptr} have been used to distinguish between
untrusted user pointers and trusted pointers in OS kernel code.
In contrast, \sys automatically applies such pointer sanitizations by
leveraging the C++ type system.

One approach to avoid double-fetch bugs is to marshal all shared data before
using it.
But, this comes at a cost.
Marshaling tools and APIs typically require array bounds annotations, which is
tedious and demands in-depth knowledge of the sandboxed library's internal data
structures.
Automatic marshaling tools like C-strider~\cite{c-strider} and
PtrSplit~\cite{ptrsplit} address this limitation; however, these tools either
impose a significant overhead or lack support for multi-threading.
\sys uses shared memory and statically enforces that shared data is placed in
shared memory, avoiding the need for custom marshaling tools or annotations.
The use of shared memory, however, introduces possible double-fetch bugs.
While \sys provides APIs to assist with double-fetches, the possibility of
unhandled double-fetch bugs still remain.
Several recent techniques detect double-fetches from shared
memory~\cite{double-fetch-static-analysis,xu2018precise, double-fetch-hardware}
and can be used alongside \sys.

Previous efforts have also sought to prevent leaking pointers that could
compromise ASLR~\cite{ASLR-guard, readactor, braden2016leakage}.
\sys prevents pointer leaks by disallowing pointers to renderer memory
to pass into the sandboxed library via the type system.

\para{Porting assistance}
Several privilege separation tools provide assistance when migrating to
a sandboxed architecture.
Wedge (Crowbar)~\cite{wedge} uses dynamic analysis to guide the
migration, and also supports an incremental porting mode that disables
isolation so that developers can test the partial port and identify the next
step.
SOAAP~\cite{soaap} uses code annotations and a custom compiler to guide
the developer.
PrivTrans~\cite{privtrans} uses code annotations and automatic source code
rewriting to separate code and data into separate components.
In contrast, \sys assists with porting without any custom tooling, purely
through the use of compile-time errors, by identifying code that must be
modified for security and shared data that must be migrated to sandbox memory
(\S\ref{sec:migration}).

\section{Using \sys in production}
\label{sec:upstream}

Over the last 6~months we've been integrating \sys into production \ff.
In this section, we describe the difference between our research prototype and
the production \sys, and our migration of the \graphite font shaping library to
use \sys.
We are in the process of migrating several other libraries and adding support
for \ff on Windows~\cite{moz-blog, ff-integration}.

\subsection{Making \sys production-ready}

To make \sys production-ready we adapt a new isolation mechanism based on
WebAssembly and rewrite the \sys API, using our prototype implementation as a
reference.

\para{SFI using WebAssembly}
In production, we use Wasm instead of NaCl to isolate library code from
the rest of \ff within a single process.
Though Wasm's performance and feature-set is not yet on par with
NaCl's~\cite{not-so-fast}, NaCl has been deprecated in favor of
Wasm~\cite{nacl-deprecate} and maintaining a separate toolchain for library
sandboxing is prohibitive.
Moreover, these limitations are likely to disappear soon: as part of the
Bytecode Alliance, multiple companies are working together to build robust Wasm
compilation toolchains, a standard syscall interface, SIMD extensions,
etc.
 
Since our goal is to reap the benefits of these efforts, we need to minimize
the changes to these toolchains.
In particular, this means that we cannot adjust for differences between the \ff
and Wasm machine model as we did for NaCl~\cite{gobi}---by intrusively
modifying the compiler, loader, runtime, etc. (\S\ref{sec:impl_sfi}).
We, instead, take advantage of the fact that \sys intercepts all data and
control flow to automatically translate between the \ff and Wasm machine models
in the \sys API.

Our only modification to the Lucet Wasm runtime is an optimized trampoline,
which we are working on upstreaming~\cite{github-zero-cost}.
Since Wasm is well-typed and has deterministic semantics, our trampolines
safely eliminate the context-and stack-switching code, reducing the cost of a
cross-boundary crossing to a function call.
This optimization was key to our shipping the sandboxed \graphite
(\S\ref{sec:migration})---it reduced the overhead of \sys by 800\%.
The details and formalization of these \emph{zero-cost trampolines} will be
presented in a separate paper.

\para{Meaningful migration error-messages}
We re-implement \sys in C++~17 and use new features---in particular
\code{if constexpr}---to customize the error messages that guide developers
during migration (\S\ref{sec:migration}).
Meaningful error messages (as opposed to a wall of generic, template failures)
is key to making \sys usable to other developers.
Although implementing custom error messages in our C++~11 prototype is possible,
it would make the implementation drastically more complex;
C++~17 allows us to keep the \sys API implementation concise (under 3K lines of
code) and give meaningful error messages.

\subsection{Isolating \graphite}
\label{sec:graphite}

We use \sys to isolate the \graphite font shaping library, creating a fresh
sandbox for each Graphite font instance.
We choose \graphite largely because the Graphite fonts are not widely used on
the web, but nevertheless \ff needs to support it for web compatibility.
This means that the library is part of \ff attack surface---and thus
memory safety bugs in \graphite are security vulnerabilities in
\ff~\cite{graphite-vulnlist-1, graphite-vulnlist-2}.

\paragraph{Evaluation}
To measure the overhead of our sandboxing, we use a micro-benchmark that
measures the page render time when reflowing text in a Graphite font ten times,
adjusting the font size each time, so font caches aren't
used.\footnote{Available at:
\url{https://jfkthame.github.io/test/udhr_urd.html}.}
We find that Wasm sandboxing imposes a 85\% overhead on the \graphite code,
which in turn slows down \ff's font rendering component (which uses \graphite
internally) by 50\%.
We attribute this slowdown largely to the nascent Wasm toolchains, which don't
yet support performance optimization on par with, say LLVM~\cite{not-so-fast,
cranelift-speedup}.
Nevertheless, this overhead is not user-perceptible; in practice page rendering
is slowed down due to the network and heavy media content, not fonts.

To measure memory overhead, we use \code{cgmemtime} to capture the peak
resident memory and cache used by \ff on the same micro-benchmark.
We find the memory overhead to be negligible---the median peak
memory overhead when loading the micro-benchmark ten times is 0.68\% (peak
memory use went from 431460~KB to 434426~KB).

\para{Deployment}
The rewritten \sys library as well as the modifications to \ff to use a sandbox 
\graphite have been merged into the \ff code base~\cite{ff-integration}.
Our retrofitted \ff successfully tested on both the \ff Nightly and Beta
channels, and ships in stock \ff~74 to Linux users and in \ff~75 to Mac
users~\cite{moz-blog}.

\section{Conclusion}

Third party libraries are likely to remain a significant source of critical
browser vulnerabilities. Our approach to sandboxing code at the
library-renderer interface offers a practical path to mitigating this threat in
\ff, and other browsers as well.

\sys shows how a type-driven approach can significantly ease the burden of
securely sandboxing libraries in existing code, through a combination of
static information flow enforcement, dynamic checks, and validations. \sys
is not dependent on \ff and is useful as a general purpose sandboxing framework
for other C++ applications.

\subsection*{Acknowledgements}
{
We thank the anonymous reviewers for their insightful feedback.
We thank our collaborators at Mozilla (especially Bobby Holley, Jonathan Kew,
Eric Rescorla, Tom Ritter, and Ricky Stewart), Fastly (especially Pat Hickey
and Tyler McMullen), and Tor (especially Georg Koppen) for fruitful discussions
and help integrating RLBox into production.
This work was supported in part by gifts from Cisco, Fastly, and Mozilla, and
by the CONIX Research Center, one of six centers in JUMP, a Semiconductor
Research Corporation (SRC) program sponsored by DARPA.
}

{
\fontsize{7}{8}\selectfont
\setlength{\bibsep}{3pt}
\bibliographystyle{abbrv}
\bibliography{local}
}

\ifExtended
\appendix
\section{Automated Pointer Swizzling}
\label{sec:appendix_swizzling}
Applications represent pointers as full 64-bit addresses (on 64-bit systems), 
however sandboxes such as the NaCl and Wasm SFI sandboxes, represents pointers 
as 32-bit offsets from the sandbox base address.
Thus pointers must be swizzled to share data structures between the application 
and sandboxed library.

\para{The \vtainted type}
To automate pointer swizzling, \sys internally adds an additional type, 
\vtainted, that it uses when dereferencing a \tainted pointer. 
Like \tainted wrappers, the \vtainted wrapper also refers to data from 
the sandboxed library, with the distinction that \vtainted data is 
stored in sandbox memory (memory accessible by sandboxed code) rather 
than application memory.
These \vtainted types are primarily used as \code{rvalue} types and 
automatically convert into \tainted types when copied into application 
memory, meaning that the programmer would never create a variable of 
\vtainted type and need not know of the type's existence.
Distinguishing between \tainted and \vtainted types allow \sys to 
automate swizzling.
By inspecting types, \sys can swizzle pointers appropriately any time 
we write \tainted pointer to a \vtainted location or vice-versa.

%

%

\para{Context free swizzling}
A further complication of implementing pointer swizzling is that \sys 
needs to swizzle pointers without knowledge of the sandbox to which 
data belongs to or the location of sandbox memory!
For instance, consider the following program. 
\begin{minted}[linenos=true, breaklines=true, mathescape=true,
escapeinside=||]{cpp}
auto mRLBox = createSandbox(...);
tainted<int**> foo = sandbox_invoke(mRLBox, ...);
tainted<int*> bar = sandbox_invoke(mRLBox, ...);
*foo = bar;|\label{line:sandbox_ptr}|
bar = *foo;|\label{line:unsandbox_ptr}|
\end{minted}
On line~\ref{line:sandbox_ptr}, we write \code{bar} to the location 
pointed to by \code{foo}.
However, as \code{foo} is a \tainted variable, \code{foo} points to a 
location in sandbox memory.
Thus dereferencing \code{foo} results in a \vtainted reference (which 
points to sandboxed memory).
So we must swizzle \code{bar} from the application pointer 
representation to the sandboxed pointer representation, prior to 
updating \code{*foo}.
The swizzling occurs in the overloaded \code{=} operator of the 
\vtainted type which in C++ has a fixed type signature and is only 
permitted to take a reference to the assigned value---it does not 
include a reference to \code{mRLBox}.
%
%
%
The reverse problem occurs on line~\ref{line:unsandbox_ptr} where the 
expression on the right is a \vtainted, while the expression on the 
left is a \tainted.
An easy albeit ugly solution could use a specific API for 
dereferencing, i.e. replacing \code{*foo} with \code{dereference(foo, 
mRLBox)}, however, such approaches changes the natural syntax expected 
by a C++ programmer.
%

\para{Implementing context free swizzling}
\sys instead implements swizzling without knowledge of the sandbox base address 
by leveraging the structure of pointer representations.
This approach is applicable for any sandbox whose base memory is 
sufficiently aligned, as is the case in our SFI implementation which is aligned 
to 4GB.
Converting from an application pointer representation to sandboxed 
pointer representation required on line~\ref{line:sandbox_ptr} can be 
easily performed by clearing the top 32-bits of pointer \code{bar} in 
the overloaded equality operator.
However, the second case which is the conversion of a sandboxed pointer 
representation to an application pointer representation on 
line~\ref{line:unsandbox_ptr} is more challenging.

To solve this, we make the observation that we can successfully learn 
the sandbox base address by inspecting the first 32-bits of any 
\emph{example pointer} which is (1) not null (2) is in the same sandbox 
as the pointer we wish to convert and (3) is stored in the application 
pointer format (as a 64-bit pointer).
The implementation of swizzling below shows how to find such an 
\emph{example pointer} below.
\begin{minted}[linenos=false, breaklines=true, mathescape=true,
escapeinside=||]{cpp}
#define top32 0xFFFFFFFF00000000
#define bot32 0xFFFFFFFF
tainted<T>& operator=(tainted_volatile<T2>& rhs) {
  uintptr_t exampleAppPtr = &(rhs.val);|\label{line:missing_exampleptr}|
  uintptr_t base = exampleAppPtr & top32;
  this->val = base + (rhs.val & bot32);
  return *this;
}
\end{minted}
Line~\ref{line:missing_exampleptr} constructs an \emph{example pointer} in the 
body of \code{operator=}, with the expression \code{&(rhs.field)}.
This expression points to the same memory location as the variable \code{foo} 
on line~\ref{line:unsandbox_ptr} in the original program.
Importantly, \code{foo} may be used as an example pointer as it is 
located in 
the same sandbox as \code{*foo}, is stored in the application pointer 
representation, and is not null as we successfully dereferenced it on 
line~\ref{line:unsandbox_ptr}, prior to the call of the \code{operator=}
also on line~\ref{line:unsandbox_ptr}.
%
%
%
%
This solution is also used to solve similar problems when supporting pointer 
arithmetic on \tainted types.
%
%
%

\section{Native Client Tooling Changes}
\label{sec:appendix_nacl}
NaCl was originally designed for sandboxing mobile code in the browser, rather 
than libraries.
We thus had to make significant modifications to the compiler, optimization 
passes, ELF loader, machine model and runtime prior to it being usable for 
library sandboxing.
We list these changes below.

\para{Loader and Runtime}
The NaCl loader does not support libraries. We thus modify the loader and 
runtime to save the ELF symbol table from when the library is loaded, to
provide symbol resolution.

\para{Threading}
On 64-bit platforms, NaCl relies on each sandbox having unique OS threads due 
to some design choices in the use of thread-local storage.  However, for 
library sandboxing, the same application thread may invoke code from different 
sandboxes; therefore we had to modify the NaCl to operate without this 
requirement.

\para{Function Invocation}
NaCl uses a ``springboard''---a piece of code which enables memory 
isolation prior to starting sandbox code---to invoke \code{main} with no 
function parameters.
For library sandboxing, we modify the springboard to allow arbitrary function 
invocations with any parameters.
We implemented callback support using the default syscall mediation mechanism 
in NaCl to correctly redirect callbacks .

\para{SIMD}
NaCl custom compilation toolchain does not support handwritten SIMD and 
assembly instructions found in libraries such as \turbo.
To mitigate this, we modify the NASM assembler to assemble this hand-written
assembly to object files that pass NaCl verifier checks.
We measured that this allowing handwritten SIMD instructions provided a 48\% 
speedup for the SFI based sandboxing on a simple benchmark rendering a \libjpeg 
image.

\para{Machine Model}
The machine model of a platform dictates the size of various types such as 
\code{int}, \code{long}, or pointers.
To maximize performance, NaCl uses a custom machine model---for 
example, NaCl uses 32-bit pointers, even in its 64-bit machine model.
However, this can introduce incompatibilities during library sandboxing when 
data structures are transferred between the application and library

To remedy this, we modified the NaCl tooling to ensure the machine model
is consistent with the \ff application code.
This required making significant changes to the NaCl Clang toolchain, the NaCl
instruction set in LLVM, NaCl's implementation of libc and the NaCl runtime.
We implemented these changes carefully such that NaCl Verifier (the tool used to
determine the safety of a NaCl binary) would not require any modifications, 
giving us confidence that our changes do not affect security.

The Wasm toolchains also exhibit similar differences in the machine model.
Rather than make relying on similar large changes in the Wasm compiler 
toolchain, we use \sys feature to automatically adjust for machine model 
differences (\S\ref{sec:upstream}) to ensure compatibility. 

\else
\fi

\end{document}